\documentclass[submission,copyright,creativecommons]{eptcs}

\providecommand{\event}{LINEARITY \& TLLA 2020} 

\renewcommand{\emph}[1]{\textbf{#1}} 

\usepackage{amsthm}
\usepackage{amsmath}
\usepackage{amssymb}
\usepackage{mdframed}
\usepackage{mathpartir}

\usepackage{todonotes}
\usepackage{supertabular}
\usepackage{hyperref}
\usepackage{rotating}
\usepackage{cmll}
\usepackage{stmaryrd}
\usepackage{thm-restate}
\usepackage{enumitem}

\usepackage{tikz-cd}

\let\Box\square
\let\square\relax

\usepackage[barr]{xy}

\let\mto\to

\newtheorem{theorem}{Theorem}

\newtheorem{definition}[theorem]{Definition}

\let\mto\to

\newcommand{\cat}[1]{\mathcal{#1}}

\newcommand{\Set}{\mathsf{Set}}
\newcommand{\PSet}{\Set_*}

\definecolor{cbsblue}{rgb}{213,94,0}

\newcommand{\ottdrule}[4][]{{\displaystyle\frac{\begin{array}{l}#2\end{array}}{#3}\quad\ottdrulename{#4}}}

\newcommand{\ottpremise}[1]{ #1 \\}
\newenvironment{ottdefnblock}[3][]{ \framebox{\mbox{#2}} \quad #3 \\[0pt]}{}

\newcommand{\ottnt}[1]{\mathit{#1}}
\newcommand{\ottmv}[1]{\mathit{#1}}

\newcommand{\ottsym}[1]{#1}

\newcommand{\ottdruleVMXXXEmptyName}[0]{\ottdrulename{VMX\_Empty}}
\newcommand{\ottdruleVMXXXEmpty}[1]{\ottdrule[#1]{%
}{
 \vdash  \ottmv{r}  \circledast   \emptyset  }{%
{\ottdruleVMXXXEmptyName}{}%
}}

\newcommand{\ottdruleVMXXXExtName}[0]{\ottdrulename{VMX\_Ext}}
\newcommand{\ottdruleVMXXXExt}[1]{\ottdrule[#1]{%
\ottpremise{  \vdash  \ottmv{r}  \circledast  \phi   \quad    \ottmv{r}  \circledast  \ottmv{r'}  \, \in \, \mathcal{R}  }%
}{
 \vdash  \ottmv{r}  \circledast   (  \phi  \ottsym{,}  \ottmv{r'}  )  }{%
{\ottdruleVMXXXExtName}{}%
}}

\newcommand{\ottdruleSGXXidName}[0]{\ottdrulename{SG\_id}}
\newcommand{\ottdruleSGXXid}[1]{\ottdrule[#1]{%
}{
  \mathsf{m}   \odot   \ottnt{X}   \vdash_{\mathcal{G} }  \ottnt{X} }{%
{\ottdruleSGXXidName}{}%
}}

\newcommand{\ottdruleSGXXUnitRName}[0]{\ottdrulename{SG\_UnitR}}
\newcommand{\ottdruleSGXXUnitR}[1]{\ottdrule[#1]{%
}{
  \emptyset   \odot   \emptyset   \vdash_{\mathcal{G} }   J  }{%
{\ottdruleSGXXUnitRName}{}%
}}

\newcommand{\ottdruleSGXXTenLName}[0]{\ottdrulename{SG\_TenL}}
\newcommand{\ottdruleSGXXTenL}[1]{\ottdrule[#1]{%
\ottpremise{  (  \phi_{{\mathrm{1}}}  \ottsym{,}  \ottmv{r}  \ottsym{,}  \ottmv{r}  \ottsym{,}  \phi_{{\mathrm{2}}}  )   \odot   ( \Phi_{{\mathrm{1}}}  \ottsym{,}   \ottnt{X}   \ottsym{,}   \ottnt{Y}   \ottsym{,}  \Phi_{{\mathrm{2}}} )   \vdash_{\mathcal{G} }  \ottnt{Z} }%
}{
  (  \phi_{{\mathrm{1}}}  \ottsym{,}  \ottmv{r}  \ottsym{,}  \phi_{{\mathrm{2}}}  )   \odot   ( \Phi_{{\mathrm{1}}}  \ottsym{,}    \ottnt{X}  \boxtimes  \ottnt{Y}    \ottsym{,}  \Phi_{{\mathrm{2}}} )   \vdash_{\mathcal{G} }  \ottnt{Z} }{%
{\ottdruleSGXXTenLName}{}%
}}

\newcommand{\ottdruleSGXXTenRName}[0]{\ottdrulename{SG\_TenR}}
\newcommand{\ottdruleSGXXTenR}[1]{\ottdrule[#1]{%
\ottpremise{ \phi_{{\mathrm{1}}}  \odot  \Phi_{{\mathrm{1}}}  \vdash_{\mathcal{G} }  \ottnt{X} }%
\ottpremise{ \phi_{{\mathrm{2}}}  \odot  \Phi_{{\mathrm{2}}}  \vdash_{\mathcal{G} }  \ottnt{Y} }%
}{
  (  \phi_{{\mathrm{1}}}  \ottsym{,}  \phi_{{\mathrm{2}}}  )   \odot   ( \Phi_{{\mathrm{1}}}  \ottsym{,}  \Phi_{{\mathrm{2}}} )   \vdash_{\mathcal{G} }   \ottnt{X}  \boxtimes  \ottnt{Y}  }{%
{\ottdruleSGXXTenRName}{}%
}}

\newcommand{\ottdruleSGXXGrdRName}[0]{\ottdrulename{SG\_GrdR}}
\newcommand{\ottdruleSGXXGrdR}[1]{\ottdrule[#1]{%
\ottpremise{ \phi  \odot  \Phi  ;   \emptyset   \vdash_{\mathcal{M} }  \ottnt{B} }%
}{
 \phi  \odot  \Phi  \vdash_{\mathcal{G} }   \mathsf{Grd}\, \ottnt{B}  }{%
{\ottdruleSGXXGrdRName}{}%
}}

\newcommand{\ottdruleSGXXCutName}[0]{\ottdrulename{SG\_Cut}}
\newcommand{\ottdruleSGXXCut}[1]{\ottdrule[#1]{%
\ottpremise{ \phi_{{\mathrm{2}}}  \odot  \Phi_{{\mathrm{2}}}  \vdash_{\mathcal{G} }  \ottnt{X} }%
\ottpremise{   (  \phi_{{\mathrm{1}}}  \ottsym{,}  \ottmv{r}  \ottsym{,}  \phi_{{\mathrm{3}}}  )   \odot   ( \Phi_{{\mathrm{1}}}  \ottsym{,}   \ottnt{X}   \ottsym{,}  \Phi_{{\mathrm{3}}} )   \vdash_{\mathcal{G} }  \ottnt{Y}   \quad   \vdash  \ottmv{r}  \circledast  \phi_{{\mathrm{2}}}  }%
}{
  (  \phi_{{\mathrm{1}}}  \ottsym{,}    \ottmv{r}  \circledast  \phi_{{\mathrm{2}}}    \ottsym{,}  \phi_{{\mathrm{3}}}  )   \odot   ( \Phi_{{\mathrm{1}}}  \ottsym{,}  \Phi_{{\mathrm{2}}}  \ottsym{,}  \Phi_{{\mathrm{3}}} )   \vdash_{\mathcal{G} }  \ottnt{Y} }{%
{\ottdruleSGXXCutName}{}%
}}

\newcommand{\ottdruleSGXXSubName}[0]{\ottdrulename{SG\_Sub}}
\newcommand{\ottdruleSGXXSub}[1]{\ottdrule[#1]{%
\ottpremise{  \phi_{{\mathrm{1}}}  \odot  \Phi_{{\mathrm{1}}}  \vdash_{\mathcal{G} }  \ottnt{X}   \quad   \phi_{{\mathrm{1}}}  \leq  \phi_{{\mathrm{2}}}  }%
}{
 \phi_{{\mathrm{2}}}  \odot  \Phi_{{\mathrm{2}}}  \vdash_{\mathcal{G} }  \ottnt{X} }{%
{\ottdruleSGXXSubName}{}%
}}

\newcommand{\ottdruleSGXXWeakName}[0]{\ottdrulename{SG\_Weak}}
\newcommand{\ottdruleSGXXWeak}[1]{\ottdrule[#1]{%
\ottpremise{   (  \phi_{{\mathrm{1}}}  \ottsym{,}  \phi_{{\mathrm{2}}}  )   \odot   ( \Phi_{{\mathrm{1}}}  \ottsym{,}  \Phi_{{\mathrm{2}}} )   \vdash_{\mathcal{G} }  \ottnt{Y}   \quad   \mathsf{a}  \, \in \, \mathcal{R} }%
}{
  (  \phi_{{\mathrm{1}}}  \ottsym{,}   \mathsf{a}   \ottsym{,}  \phi_{{\mathrm{2}}}  )   \odot   ( \Phi_{{\mathrm{1}}}  \ottsym{,}   \ottnt{X}   \ottsym{,}  \Phi_{{\mathrm{2}}} )   \vdash_{\mathcal{G} }  \ottnt{Y} }{%
{\ottdruleSGXXWeakName}{}%
}}

\newcommand{\ottdruleSGXXContName}[0]{\ottdrulename{SG\_Cont}}
\newcommand{\ottdruleSGXXCont}[1]{\ottdrule[#1]{%
\ottpremise{   (  \phi_{{\mathrm{1}}}  \ottsym{,}  \ottmv{r_{{\mathrm{1}}}}  \ottsym{,}  \ottmv{r_{{\mathrm{2}}}}  \ottsym{,}  \phi_{{\mathrm{2}}}  )   \odot   ( \Phi_{{\mathrm{1}}}  \ottsym{,}   \ottnt{X}   \ottsym{,}   \ottnt{X}   \ottsym{,}  \Phi_{{\mathrm{2}}} )   \vdash_{\mathcal{G} }  \ottnt{Y}   \quad  \ottsym{(}   \ottmv{r_{{\mathrm{1}}}}  \oplus  \ottmv{r_{{\mathrm{2}}}}   \ottsym{)} \, \in \, \mathcal{R} }%
}{
  (  \phi_{{\mathrm{1}}}  \ottsym{,}   \ottmv{r_{{\mathrm{1}}}}  \oplus  \ottmv{r_{{\mathrm{2}}}}   \ottsym{,}  \phi_{{\mathrm{2}}}  )   \odot   ( \Phi_{{\mathrm{1}}}  \ottsym{,}   \ottnt{X}   \ottsym{,}  \Phi_{{\mathrm{2}}} )   \vdash_{\mathcal{G} }  \ottnt{Y} }{%
{\ottdruleSGXXContName}{}%
}}

\newcommand{\ottdruleSGXXExName}[0]{\ottdrulename{SG\_Ex}}
\newcommand{\ottdruleSGXXEx}[1]{\ottdrule[#1]{%
\ottpremise{  (  \phi_{{\mathrm{1}}}  \ottsym{,}  \ottmv{r_{{\mathrm{1}}}}  \ottsym{,}  \ottmv{r_{{\mathrm{2}}}}  \ottsym{,}  \phi_{{\mathrm{2}}}  )   \odot   ( \Phi_{{\mathrm{1}}}  \ottsym{,}   \ottnt{X}   \ottsym{,}   \ottnt{Y}   \ottsym{,}  \Phi_{{\mathrm{2}}} )   \vdash_{\mathcal{G} }  \ottnt{Z} }%
}{
  (  \phi_{{\mathrm{1}}}  \ottsym{,}  \ottmv{r_{{\mathrm{2}}}}  \ottsym{,}  \ottmv{r_{{\mathrm{1}}}}  \ottsym{,}  \phi_{{\mathrm{2}}}  )   \odot   ( \Phi_{{\mathrm{1}}}  \ottsym{,}   \ottnt{Y}   \ottsym{,}   \ottnt{X}   \ottsym{,}  \Phi_{{\mathrm{2}}} )   \vdash_{\mathcal{G} }  \ottnt{Z} }{%
{\ottdruleSGXXExName}{}%
}}

\newcommand{\ottdruleSMXXIdName}[0]{\ottdrulename{SM\_Id}}
\newcommand{\ottdruleSMXXId}[1]{\ottdrule[#1]{%
}{
  \emptyset   \odot   \emptyset   ;  \ottnt{A}  \vdash_{\mathcal{M} }  \ottnt{A} }{%
{\ottdruleSMXXIdName}{}%
}}

\newcommand{\ottdruleSMXXUnitRName}[0]{\ottdrulename{SM\_UnitR}}
\newcommand{\ottdruleSMXXUnitR}[1]{\ottdrule[#1]{%
}{
  \emptyset   \odot   \emptyset   ;   \emptyset   \vdash_{\mathcal{M} }   I  }{%
{\ottdruleSMXXUnitRName}{}%
}}

\newcommand{\ottdruleSMXXGTenLName}[0]{\ottdrulename{SM\_GTenL}}
\newcommand{\ottdruleSMXXGTenL}[1]{\ottdrule[#1]{%
\ottpremise{  (  \phi_{{\mathrm{1}}}  \ottsym{,}  \ottmv{r}  \ottsym{,}  \ottmv{r}  \ottsym{,}  \phi_{{\mathrm{2}}}  )   \odot   ( \Phi_{{\mathrm{1}}}  \ottsym{,}   \ottnt{X}   \ottsym{,}   \ottnt{Y}   \ottsym{,}  \Phi_{{\mathrm{2}}} )   ;  \Gamma  \vdash_{\mathcal{M} }  \ottnt{B} }%
}{
  (  \phi_{{\mathrm{1}}}  \ottsym{,}  \ottmv{r}  \ottsym{,}  \phi_{{\mathrm{2}}}  )   \odot   ( \Phi_{{\mathrm{1}}}  \ottsym{,}    \ottnt{X}  \boxtimes  \ottnt{Y}    \ottsym{,}  \Phi_{{\mathrm{2}}} )   ;  \Gamma  \vdash_{\mathcal{M} }  \ottnt{B} }{%
{\ottdruleSMXXGTenLName}{}%
}}

\newcommand{\ottdruleSMXXTenLName}[0]{\ottdrulename{SM\_TenL}}
\newcommand{\ottdruleSMXXTenL}[1]{\ottdrule[#1]{%
\ottpremise{ \phi  \odot  \Phi  ;   ( \Gamma_{{\mathrm{1}}}  \ottsym{,}  \ottnt{A}  \ottsym{,}  \ottnt{B}  \ottsym{,}  \Gamma_{{\mathrm{2}}} )   \vdash_{\mathcal{M} }  \ottnt{C} }%
}{
 \phi  \odot  \Phi  ;   ( \Gamma_{{\mathrm{1}}}  \ottsym{,}   \ottnt{A}  \otimes  \ottnt{B}   \ottsym{,}  \Gamma_{{\mathrm{2}}} )   \vdash_{\mathcal{M} }  \ottnt{C} }{%
{\ottdruleSMXXTenLName}{}%
}}

\newcommand{\ottdruleSMXXTenRName}[0]{\ottdrulename{SM\_TenR}}
\newcommand{\ottdruleSMXXTenR}[1]{\ottdrule[#1]{%
\ottpremise{ \phi_{{\mathrm{1}}}  \odot  \Phi_{{\mathrm{1}}}  ;  \Gamma_{{\mathrm{1}}}  \vdash_{\mathcal{M} }  \ottnt{A} }%
\ottpremise{ \phi_{{\mathrm{2}}}  \odot  \Phi_{{\mathrm{2}}}  ;  \Gamma_{{\mathrm{2}}}  \vdash_{\mathcal{M} }  \ottnt{B} }%
}{
  (  \phi_{{\mathrm{1}}}  \ottsym{,}  \phi_{{\mathrm{2}}}  )   \odot   ( \Phi_{{\mathrm{1}}}  \ottsym{,}  \Phi_{{\mathrm{2}}} )   ;   ( \Gamma_{{\mathrm{1}}}  \ottsym{,}  \Gamma_{{\mathrm{2}}} )   \vdash_{\mathcal{M} }   \ottnt{A}  \otimes  \ottnt{B}  }{%
{\ottdruleSMXXTenRName}{}%
}}

\newcommand{\ottdruleSMXXImpLName}[0]{\ottdrulename{SM\_ImpL}}
\newcommand{\ottdruleSMXXImpL}[1]{\ottdrule[#1]{%
\ottpremise{ \phi_{{\mathrm{2}}}  \odot  \Phi_{{\mathrm{2}}}  ;  \Gamma_{{\mathrm{2}}}  \vdash_{\mathcal{M} }  \ottnt{A} }%
\ottpremise{ \phi_{{\mathrm{1}}}  \odot  \Phi_{{\mathrm{1}}}  ;   ( \Gamma_{{\mathrm{1}}}  \ottsym{,}  \ottnt{B}  \ottsym{,}  \Gamma_{{\mathrm{3}}} )   \vdash_{\mathcal{M} }  \ottnt{C} }%
}{
  (  \phi_{{\mathrm{1}}}  \ottsym{,}  \phi_{{\mathrm{2}}}  )   \odot   ( \Phi_{{\mathrm{1}}}  \ottsym{,}  \Phi_{{\mathrm{2}}} )   ;   ( \Gamma_{{\mathrm{1}}}  \ottsym{,}  \ottsym{(}   \ottnt{A}  \multimap  \ottnt{B}   \ottsym{)}  \ottsym{,}  \Gamma_{{\mathrm{2}}}  \ottsym{,}  \Gamma_{{\mathrm{3}}} )   \vdash_{\mathcal{M} }  \ottnt{C} }{%
{\ottdruleSMXXImpLName}{}%
}}

\newcommand{\ottdruleSMXXImpRName}[0]{\ottdrulename{SM\_ImpR}}
\newcommand{\ottdruleSMXXImpR}[1]{\ottdrule[#1]{%
\ottpremise{ \phi  \odot  \Phi  ;   ( \Gamma  \ottsym{,}  \ottnt{A} )   \vdash_{\mathcal{M} }  \ottnt{B} }%
}{
 \phi  \odot  \Phi  ;  \Gamma  \vdash_{\mathcal{M} }   \ottnt{A}  \multimap  \ottnt{B}  }{%
{\ottdruleSMXXImpRName}{}%
}}

\newcommand{\ottdruleSMXXGrdLName}[0]{\ottdrulename{SM\_GrdL}}
\newcommand{\ottdruleSMXXGrdL}[1]{\ottdrule[#1]{%
\ottpremise{ \phi  \odot  \Phi  ;   ( \ottnt{A}  \ottsym{,}  \Gamma )   \vdash_{\mathcal{M} }  \ottnt{B} }%
}{
  (  \phi  \ottsym{,}   \mathsf{m}   )   \odot   ( \Phi  \ottsym{,}    \mathsf{Grd}\, \ottnt{A}   )   ;  \Gamma  \vdash_{\mathcal{M} }  \ottnt{B} }{%
{\ottdruleSMXXGrdLName}{}%
}}

\newcommand{\ottdruleSMXXLinLName}[0]{\ottdrulename{SM\_LinL}}
\newcommand{\ottdruleSMXXLinL}[1]{\ottdrule[#1]{%
\ottpremise{  (  \phi  \ottsym{,}  \ottmv{r}  )   \odot   ( \Phi  \ottsym{,}   \ottnt{X}  )   ;  \Gamma  \vdash_{\mathcal{M} }  \ottnt{C} }%
}{
 \phi  \odot  \Phi  ;   (  \mathsf{Lin}_{ \ottmv{r} }\, \ottnt{X}   \ottsym{,}  \Gamma )   \vdash_{\mathcal{M} }  \ottnt{C} }{%
{\ottdruleSMXXLinLName}{}%
}}

\newcommand{\ottdruleSMXXLinRName}[0]{\ottdrulename{SM\_LinR}}
\newcommand{\ottdruleSMXXLinR}[1]{\ottdrule[#1]{%
\ottpremise{  \phi  \odot  \Phi  \vdash_{\mathcal{G} }  \ottnt{X}   \quad   \vdash  \ottmv{r}  \circledast  \phi  }%
}{
  (   \ottmv{r}  \circledast  \phi   )   \odot  \Phi  ;   \emptyset   \vdash_{\mathcal{M} }   \mathsf{Lin}_{ \ottmv{r} }\, \ottnt{X}  }{%
{\ottdruleSMXXLinRName}{}%
}}

\newcommand{\ottdruleSMXXCutName}[0]{\ottdrulename{SM\_Cut}}
\newcommand{\ottdruleSMXXCut}[1]{\ottdrule[#1]{%
\ottpremise{ \phi_{{\mathrm{2}}}  \odot  \Phi_{{\mathrm{2}}}  ;  \Gamma_{{\mathrm{2}}}  \vdash_{\mathcal{M} }  \ottnt{A} }%
\ottpremise{ \phi_{{\mathrm{2}}}  \odot  \Phi_{{\mathrm{1}}}  ;   ( \Gamma_{{\mathrm{1}}}  \ottsym{,}  \ottnt{A}  \ottsym{,}  \Gamma_{{\mathrm{3}}} )   \vdash_{\mathcal{M} }  \ottnt{B} }%
}{
  (  \phi_{{\mathrm{1}}}  \ottsym{,}  \phi_{{\mathrm{2}}}  )   \odot   ( \Phi_{{\mathrm{1}}}  \ottsym{,}  \Phi_{{\mathrm{2}}} )   ;   ( \Gamma_{{\mathrm{1}}}  \ottsym{,}  \Gamma_{{\mathrm{2}}}  \ottsym{,}  \Gamma_{{\mathrm{3}}} )   \vdash_{\mathcal{M} }  \ottnt{B} }{%
{\ottdruleSMXXCutName}{}%
}}

\newcommand{\ottdruleSMXXGCutName}[0]{\ottdrulename{SM\_GCut}}
\newcommand{\ottdruleSMXXGCut}[1]{\ottdrule[#1]{%
\ottpremise{ \phi_{{\mathrm{2}}}  \odot  \Phi_{{\mathrm{2}}}  \vdash_{\mathcal{G} }  \ottnt{X} }%
\ottpremise{   (  \phi_{{\mathrm{1}}}  \ottsym{,}  \ottmv{r}  \ottsym{,}  \phi_{{\mathrm{3}}}  )   \odot   ( \Phi_{{\mathrm{1}}}  \ottsym{,}   \ottnt{X}   \ottsym{,}  \Phi_{{\mathrm{3}}} )   ;  \Gamma  \vdash_{\mathcal{M} }  \ottnt{B}   \quad   \vdash  \ottmv{r}  \circledast  \phi_{{\mathrm{2}}}  }%
}{
  (  \phi_{{\mathrm{1}}}  \ottsym{,}    \ottmv{r}  \circledast  \phi_{{\mathrm{2}}}    \ottsym{,}  \phi_{{\mathrm{3}}}  )   \odot   ( \Phi_{{\mathrm{1}}}  \ottsym{,}  \Phi_{{\mathrm{2}}}  \ottsym{,}  \Phi_{{\mathrm{3}}} )   ;  \Gamma  \vdash_{\mathcal{M} }  \ottnt{B} }{%
{\ottdruleSMXXGCutName}{}%
}}

\newcommand{\ottdruleSMXXGSubName}[0]{\ottdrulename{SM\_GSub}}
\newcommand{\ottdruleSMXXGSub}[1]{\ottdrule[#1]{%
\ottpremise{  \phi_{{\mathrm{1}}}  \odot  \Phi_{{\mathrm{1}}}  ;  \Gamma  \vdash_{\mathcal{M} }  \ottnt{B}   \quad   \phi_{{\mathrm{1}}}  \leq  \phi_{{\mathrm{2}}}  }%
}{
 \phi_{{\mathrm{2}}}  \odot  \Phi_{{\mathrm{2}}}  ;  \Gamma  \vdash_{\mathcal{M} }  \ottnt{B} }{%
{\ottdruleSMXXGSubName}{}%
}}

\newcommand{\ottdruleSMXXGWeakName}[0]{\ottdrulename{SM\_GWeak}}
\newcommand{\ottdruleSMXXGWeak}[1]{\ottdrule[#1]{%
\ottpremise{   (  \phi_{{\mathrm{1}}}  \ottsym{,}  \phi_{{\mathrm{2}}}  )   \odot   ( \Phi_{{\mathrm{1}}}  \ottsym{,}  \Phi_{{\mathrm{2}}} )   ;  \Gamma  \vdash_{\mathcal{M} }  \ottnt{B}   \quad   \mathsf{a}  \, \in \, \mathcal{R} }%
}{
  (  \phi_{{\mathrm{1}}}  \ottsym{,}   \mathsf{a}   \ottsym{,}  \phi_{{\mathrm{2}}}  )   \odot   ( \Phi_{{\mathrm{1}}}  \ottsym{,}   \ottnt{X}   \ottsym{,}  \Phi_{{\mathrm{2}}} )   ;  \Gamma  \vdash_{\mathcal{M} }  \ottnt{B} }{%
{\ottdruleSMXXGWeakName}{}%
}}

\newcommand{\ottdruleSMXXGContName}[0]{\ottdrulename{SM\_GCont}}
\newcommand{\ottdruleSMXXGCont}[1]{\ottdrule[#1]{%
\ottpremise{   (  \phi_{{\mathrm{1}}}  \ottsym{,}  \ottmv{r_{{\mathrm{1}}}}  \ottsym{,}  \ottmv{r_{{\mathrm{2}}}}  \ottsym{,}  \phi_{{\mathrm{2}}}  )   \odot   ( \Phi_{{\mathrm{1}}}  \ottsym{,}   \ottnt{X}   \ottsym{,}   \ottnt{X}   \ottsym{,}  \Phi_{{\mathrm{2}}} )   ;  \Gamma  \vdash_{\mathcal{M} }  \ottnt{B}   \quad  \ottsym{(}   \ottmv{r_{{\mathrm{1}}}}  \oplus  \ottmv{r_{{\mathrm{2}}}}   \ottsym{)} \, \in \, \mathcal{R} }%
}{
  (  \phi_{{\mathrm{1}}}  \ottsym{,}   \ottmv{r_{{\mathrm{1}}}}  \oplus  \ottmv{r_{{\mathrm{2}}}}   \ottsym{,}  \phi_{{\mathrm{2}}}  )   \odot   ( \Phi_{{\mathrm{1}}}  \ottsym{,}   \ottnt{X}   \ottsym{,}  \Phi_{{\mathrm{2}}} )   ;  \Gamma  \vdash_{\mathcal{M} }  \ottnt{B} }{%
{\ottdruleSMXXGContName}{}%
}}

\newcommand{\ottdruleSMXXGExName}[0]{\ottdrulename{SM\_GEx}}
\newcommand{\ottdruleSMXXGEx}[1]{\ottdrule[#1]{%
\ottpremise{  (  \phi_{{\mathrm{1}}}  \ottsym{,}  \ottmv{r_{{\mathrm{1}}}}  \ottsym{,}  \ottmv{r_{{\mathrm{2}}}}  \ottsym{,}  \phi_{{\mathrm{2}}}  )   \odot   ( \Phi_{{\mathrm{1}}}  \ottsym{,}   \ottnt{X}   \ottsym{,}   \ottnt{Y}   \ottsym{,}  \Phi_{{\mathrm{2}}} )   ;  \Gamma  \vdash_{\mathcal{M} }  \ottnt{B} }%
}{
  (  \phi_{{\mathrm{1}}}  \ottsym{,}  \ottmv{r_{{\mathrm{2}}}}  \ottsym{,}  \ottmv{r_{{\mathrm{1}}}}  \ottsym{,}  \phi_{{\mathrm{2}}}  )   \odot   ( \Phi_{{\mathrm{1}}}  \ottsym{,}   \ottnt{Y}   \ottsym{,}   \ottnt{X}   \ottsym{,}  \Phi_{{\mathrm{2}}} )   ;  \Gamma  \vdash_{\mathcal{M} }  \ottnt{B} }{%
{\ottdruleSMXXGExName}{}%
}}

\newcommand{\ottdruleSMXXExName}[0]{\ottdrulename{SM\_Ex}}
\newcommand{\ottdruleSMXXEx}[1]{\ottdrule[#1]{%
\ottpremise{ \phi  \odot  \Phi  ;   ( \Gamma_{{\mathrm{1}}}  \ottsym{,}  \ottnt{A}  \ottsym{,}  \ottnt{B}  \ottsym{,}  \Gamma_{{\mathrm{2}}} )   \vdash_{\mathcal{M} }  \ottnt{B} }%
}{
 \phi  \odot  \Phi  ;   ( \Gamma_{{\mathrm{1}}}  \ottsym{,}  \ottnt{B}  \ottsym{,}  \ottnt{A}  \ottsym{,}  \Gamma_{{\mathrm{2}}} )   \vdash_{\mathcal{M} }  \ottnt{B} }{%
{\ottdruleSMXXExName}{}%
}}

\newcommand{\ottdruleNGXXIdName}[0]{\ottdrulename{NG\_Id}}
\newcommand{\ottdruleNGXXId}[1]{\ottdrule[#1]{%
}{
  \mathsf{m}   \odot   \ottnt{X}   \vdash_{\mathcal{G} }  \ottnt{X} }{%
{\ottdruleNGXXIdName}{}%
}}

\newcommand{\ottdruleNGXXUnitIName}[0]{\ottdrulename{NG\_UnitI}}
\newcommand{\ottdruleNGXXUnitI}[1]{\ottdrule[#1]{%
}{
  \emptyset   \odot   \emptyset   \vdash_{\mathcal{G} }   J  }{%
{\ottdruleNGXXUnitIName}{}%
}}

\newcommand{\ottdruleNGXXTenIName}[0]{\ottdrulename{NG\_TenI}}
\newcommand{\ottdruleNGXXTenI}[1]{\ottdrule[#1]{%
\ottpremise{ \phi_{{\mathrm{1}}}  \odot  \Phi_{{\mathrm{1}}}  \vdash_{\mathcal{G} }  \ottnt{X} }%
\ottpremise{ \phi_{{\mathrm{2}}}  \odot  \Phi_{{\mathrm{2}}}  \vdash_{\mathcal{G} }  \ottnt{Y} }%
}{
  (  \phi_{{\mathrm{1}}}  \ottsym{,}  \phi_{{\mathrm{2}}}  )   \odot   ( \Phi_{{\mathrm{1}}}  \ottsym{,}  \Phi_{{\mathrm{2}}} )   \vdash_{\mathcal{G} }   \ottnt{X}  \boxtimes  \ottnt{Y}  }{%
{\ottdruleNGXXTenIName}{}%
}}

\newcommand{\ottdruleNGXXTenEName}[0]{\ottdrulename{NG\_TenE}}
\newcommand{\ottdruleNGXXTenE}[1]{\ottdrule[#1]{%
\ottpremise{ \phi_{{\mathrm{2}}}  \odot  \Phi_{{\mathrm{2}}}  \vdash_{\mathcal{G} }   \ottnt{X}  \boxtimes  \ottnt{Y}  }%
\ottpremise{   (  \phi_{{\mathrm{1}}}  \ottsym{,}  \ottmv{r}  \ottsym{,}  \ottmv{r}  \ottsym{,}  \phi_{{\mathrm{3}}}  )   \odot   ( \Phi_{{\mathrm{1}}}  \ottsym{,}   \ottnt{X}   \ottsym{,}   \ottnt{Y}   \ottsym{,}  \Phi_{{\mathrm{3}}} )   \vdash_{\mathcal{G} }  \ottnt{Z}   \quad   \vdash  \ottmv{r}  \circledast  \phi_{{\mathrm{2}}}  }%
}{
  (  \phi_{{\mathrm{1}}}  \ottsym{,}    \ottmv{r}  \circledast  \phi_{{\mathrm{2}}}    \ottsym{,}  \phi_{{\mathrm{3}}}  )   \odot   ( \Phi_{{\mathrm{1}}}  \ottsym{,}  \Phi_{{\mathrm{2}}}  \ottsym{,}  \Phi_{{\mathrm{3}}} )   \vdash_{\mathcal{G} }  \ottnt{Z} }{%
{\ottdruleNGXXTenEName}{}%
}}

\newcommand{\ottdruleNGXXGrdIName}[0]{\ottdrulename{NG\_GrdI}}
\newcommand{\ottdruleNGXXGrdI}[1]{\ottdrule[#1]{%
\ottpremise{ \phi  \odot  \Phi  ;   \emptyset   \vdash_{\mathcal{M} }  \ottnt{B} }%
}{
 \phi  \odot  \Phi  \vdash_{\mathcal{G} }   \mathsf{Grd}\, \ottnt{B}  }{%
{\ottdruleNGXXGrdIName}{}%
}}

\newcommand{\ottdruleNGXXSubName}[0]{\ottdrulename{NG\_Sub}}
\newcommand{\ottdruleNGXXSub}[1]{\ottdrule[#1]{%
\ottpremise{  \phi_{{\mathrm{1}}}  \odot  \Phi_{{\mathrm{1}}}  \vdash_{\mathcal{G} }  \ottnt{X}   \quad   \phi_{{\mathrm{1}}}  \leq  \phi_{{\mathrm{2}}}  }%
}{
 \phi_{{\mathrm{2}}}  \odot  \Phi_{{\mathrm{2}}}  \vdash_{\mathcal{G} }  \ottnt{X} }{%
{\ottdruleNGXXSubName}{}%
}}

\newcommand{\ottdruleNGXXWeakName}[0]{\ottdrulename{NG\_Weak}}
\newcommand{\ottdruleNGXXWeak}[1]{\ottdrule[#1]{%
\ottpremise{   (  \phi_{{\mathrm{1}}}  \ottsym{,}  \phi_{{\mathrm{2}}}  )   \odot   ( \Phi_{{\mathrm{1}}}  \ottsym{,}  \Phi_{{\mathrm{2}}} )   \vdash_{\mathcal{G} }  \ottnt{Y}   \quad   \mathsf{a}  \, \in \, \mathcal{R} }%
}{
  (  \phi_{{\mathrm{1}}}  \ottsym{,}   \mathsf{a}   \ottsym{,}  \phi_{{\mathrm{2}}}  )   \odot   ( \Phi_{{\mathrm{1}}}  \ottsym{,}   \ottnt{X}   \ottsym{,}  \Phi_{{\mathrm{2}}} )   \vdash_{\mathcal{G} }  \ottnt{Y} }{%
{\ottdruleNGXXWeakName}{}%
}}

\newcommand{\ottdruleNGXXContName}[0]{\ottdrulename{NG\_Cont}}
\newcommand{\ottdruleNGXXCont}[1]{\ottdrule[#1]{%
\ottpremise{   (  \phi_{{\mathrm{1}}}  \ottsym{,}  \ottmv{r_{{\mathrm{1}}}}  \ottsym{,}  \ottmv{r_{{\mathrm{2}}}}  \ottsym{,}  \phi_{{\mathrm{2}}}  )   \odot   ( \Phi_{{\mathrm{1}}}  \ottsym{,}   \ottnt{X}   \ottsym{,}   \ottnt{X}   \ottsym{,}  \Phi_{{\mathrm{2}}} )   \vdash_{\mathcal{G} }  \ottnt{Y}   \quad  \ottsym{(}   \ottmv{r_{{\mathrm{1}}}}  \oplus  \ottmv{r_{{\mathrm{2}}}}   \ottsym{)} \, \in \, \mathcal{R} }%
}{
  (  \phi_{{\mathrm{1}}}  \ottsym{,}   \ottmv{r_{{\mathrm{1}}}}  \oplus  \ottmv{r_{{\mathrm{2}}}}   \ottsym{,}  \phi_{{\mathrm{2}}}  )   \odot   ( \Phi_{{\mathrm{1}}}  \ottsym{,}   \ottnt{X}   \ottsym{,}  \Phi_{{\mathrm{2}}} )   \vdash_{\mathcal{G} }  \ottnt{Y} }{%
{\ottdruleNGXXContName}{}%
}}

\newcommand{\ottdruleNGXXExName}[0]{\ottdrulename{NG\_Ex}}
\newcommand{\ottdruleNGXXEx}[1]{\ottdrule[#1]{%
\ottpremise{  (  \phi_{{\mathrm{1}}}  \ottsym{,}  \ottmv{r_{{\mathrm{1}}}}  \ottsym{,}  \ottmv{r_{{\mathrm{2}}}}  \ottsym{,}  \phi_{{\mathrm{2}}}  )   \odot   ( \Phi_{{\mathrm{1}}}  \ottsym{,}   \ottnt{X}   \ottsym{,}   \ottnt{Y}   \ottsym{,}  \Phi_{{\mathrm{2}}} )   \vdash_{\mathcal{G} }  \ottnt{Z} }%
}{
  (  \phi_{{\mathrm{1}}}  \ottsym{,}  \ottmv{r_{{\mathrm{2}}}}  \ottsym{,}  \ottmv{r_{{\mathrm{1}}}}  \ottsym{,}  \phi_{{\mathrm{2}}}  )   \odot   ( \Phi_{{\mathrm{1}}}  \ottsym{,}   \ottnt{Y}   \ottsym{,}   \ottnt{X}   \ottsym{,}  \Phi_{{\mathrm{2}}} )   \vdash_{\mathcal{G} }  \ottnt{Z} }{%
{\ottdruleNGXXExName}{}%
}}

\newcommand{\ottdruleNMXXIdName}[0]{\ottdrulename{NM\_Id}}
\newcommand{\ottdruleNMXXId}[1]{\ottdrule[#1]{%
}{
  \emptyset   \odot   \emptyset   ;  \ottnt{A}  \vdash_{\mathcal{M} }  \ottnt{A} }{%
{\ottdruleNMXXIdName}{}%
}}

\newcommand{\ottdruleNMXXUnitIName}[0]{\ottdrulename{NM\_UnitI}}
\newcommand{\ottdruleNMXXUnitI}[1]{\ottdrule[#1]{%
}{
  \emptyset   \odot   \emptyset   ;   \emptyset   \vdash_{\mathcal{M} }   I  }{%
{\ottdruleNMXXUnitIName}{}%
}}

\newcommand{\ottdruleNMXXTenIName}[0]{\ottdrulename{NM\_TenI}}
\newcommand{\ottdruleNMXXTenI}[1]{\ottdrule[#1]{%
\ottpremise{ \phi_{{\mathrm{1}}}  \odot  \Phi_{{\mathrm{1}}}  ;  \Gamma_{{\mathrm{1}}}  \vdash_{\mathcal{M} }  \ottnt{A} }%
\ottpremise{ \phi_{{\mathrm{2}}}  \odot  \Phi_{{\mathrm{2}}}  ;  \Gamma_{{\mathrm{2}}}  \vdash_{\mathcal{M} }  \ottnt{B} }%
}{
  (  \phi_{{\mathrm{1}}}  \ottsym{,}  \phi_{{\mathrm{2}}}  )   \odot   ( \Phi_{{\mathrm{1}}}  \ottsym{,}  \Phi_{{\mathrm{2}}} )   ;   ( \Gamma_{{\mathrm{1}}}  \ottsym{,}  \Gamma_{{\mathrm{2}}} )   \vdash_{\mathcal{M} }   \ottnt{A}  \otimes  \ottnt{B}  }{%
{\ottdruleNMXXTenIName}{}%
}}

\newcommand{\ottdruleNMXXTenEName}[0]{\ottdrulename{NM\_TenE}}
\newcommand{\ottdruleNMXXTenE}[1]{\ottdrule[#1]{%
\ottpremise{ \phi_{{\mathrm{2}}}  \odot  \Phi_{{\mathrm{2}}}  ;  \Gamma_{{\mathrm{2}}}  \vdash_{\mathcal{M} }   \ottnt{A}  \otimes  \ottnt{B}  }%
\ottpremise{ \phi_{{\mathrm{1}}}  \odot  \Phi_{{\mathrm{1}}}  ;  \Gamma_{{\mathrm{1}}}  \ottsym{,}  \ottnt{A}  \ottsym{,}  \ottnt{B}  \ottsym{,}  \Gamma_{{\mathrm{3}}}  \vdash_{\mathcal{M} }  \ottnt{C} }%
}{
  (  \phi_{{\mathrm{1}}}  \ottsym{,}  \phi_{{\mathrm{2}}}  )   \odot   ( \Phi_{{\mathrm{1}}}  \ottsym{,}  \Phi_{{\mathrm{2}}} )   ;   ( \Gamma_{{\mathrm{1}}}  \ottsym{,}  \Gamma_{{\mathrm{2}}}  \ottsym{,}  \Gamma_{{\mathrm{3}}} )   \vdash_{\mathcal{M} }  \ottnt{C} }{%
{\ottdruleNMXXTenEName}{}%
}}

\newcommand{\ottdruleNMXXImpIName}[0]{\ottdrulename{NM\_ImpI}}
\newcommand{\ottdruleNMXXImpI}[1]{\ottdrule[#1]{%
\ottpremise{ \phi  \odot  \Phi  ;   ( \Gamma  \ottsym{,}  \ottnt{A} )   \vdash_{\mathcal{M} }  \ottnt{B} }%
}{
 \phi  \odot  \Phi  ;  \Gamma  \vdash_{\mathcal{M} }   \ottnt{A}  \multimap  \ottnt{B}  }{%
{\ottdruleNMXXImpIName}{}%
}}

\newcommand{\ottdruleNMXXImpEName}[0]{\ottdrulename{NM\_ImpE}}
\newcommand{\ottdruleNMXXImpE}[1]{\ottdrule[#1]{%
\ottpremise{ \phi_{{\mathrm{2}}}  \odot  \Phi_{{\mathrm{2}}}  ;  \Gamma_{{\mathrm{2}}}  \vdash_{\mathcal{M} }  \ottnt{A} }%
\ottpremise{ \phi_{{\mathrm{1}}}  \odot  \Phi_{{\mathrm{1}}}  ;  \Gamma_{{\mathrm{1}}}  \vdash_{\mathcal{M} }   \ottnt{A}  \multimap  \ottnt{B}  }%
}{
  (  \phi_{{\mathrm{1}}}  \ottsym{,}  \phi_{{\mathrm{2}}}  )   \odot   ( \Phi_{{\mathrm{1}}}  \ottsym{,}  \Phi_{{\mathrm{2}}} )   ;   ( \Gamma_{{\mathrm{1}}}  \ottsym{,}  \Gamma_{{\mathrm{2}}} )   \vdash_{\mathcal{M} }  \ottnt{B} }{%
{\ottdruleNMXXImpEName}{}%
}}

\newcommand{\ottdruleNMXXLinIName}[0]{\ottdrulename{NM\_LinI}}
\newcommand{\ottdruleNMXXLinI}[1]{\ottdrule[#1]{%
\ottpremise{  \phi  \odot  \Phi  \vdash_{\mathcal{G} }  \ottnt{X}   \quad   \vdash  \ottmv{r}  \circledast  \phi  }%
}{
  \ottmv{r}  \circledast  \phi   \odot  \Phi  ;   \emptyset   \vdash_{\mathcal{M} }   \mathsf{Lin}_{ \ottmv{r} }\, \ottnt{X}  }{%
{\ottdruleNMXXLinIName}{}%
}}

\newcommand{\ottdruleNMXXLinEName}[0]{\ottdrulename{NM\_LinE}}
\newcommand{\ottdruleNMXXLinE}[1]{\ottdrule[#1]{%
\ottpremise{ \phi_{{\mathrm{2}}}  \odot  \Phi_{{\mathrm{2}}}  ;  \Gamma_{{\mathrm{2}}}  \vdash_{\mathcal{M} }   \mathsf{Lin}_{ \ottmv{r} }\, \ottnt{X}  }%
\ottpremise{  (  \phi_{{\mathrm{1}}}  \ottsym{,}  \ottmv{r}  \ottsym{,}  \phi_{{\mathrm{2}}}  )   \odot   ( \Phi_{{\mathrm{1}}}  \ottsym{,}   \ottnt{X}   \ottsym{,}  \Phi_{{\mathrm{3}}} )   ;  \Gamma_{{\mathrm{1}}}  \vdash_{\mathcal{M} }  \ottnt{B} }%
}{
  (  \phi_{{\mathrm{1}}}  \ottsym{,}  \phi_{{\mathrm{2}}}  \ottsym{,}  \phi_{{\mathrm{3}}}  )   \odot   ( \Phi_{{\mathrm{1}}}  \ottsym{,}  \Phi_{{\mathrm{2}}}  \ottsym{,}  \Phi_{{\mathrm{3}}} )   ;   ( \Gamma_{{\mathrm{1}}}  \ottsym{,}  \Gamma_{{\mathrm{2}}} )   \vdash_{\mathcal{M} }  \ottnt{B} }{%
{\ottdruleNMXXLinEName}{}%
}}

\newcommand{\ottdruleNMXXGrdEName}[0]{\ottdrulename{NM\_GrdE}}
\newcommand{\ottdruleNMXXGrdE}[1]{\ottdrule[#1]{%
\ottpremise{ \phi  \odot  \Phi  \vdash_{\mathcal{G} }   \mathsf{Grd}\, \ottnt{A}  }%
}{
 \phi  \odot  \Phi  ;   \emptyset   \vdash_{\mathcal{M} }  \ottnt{A} }{%
{\ottdruleNMXXGrdEName}{}%
}}

\newcommand{\ottdruleNMXXGSubName}[0]{\ottdrulename{NM\_GSub}}
\newcommand{\ottdruleNMXXGSub}[1]{\ottdrule[#1]{%
\ottpremise{  \phi_{{\mathrm{1}}}  \odot  \Phi  ;  \Gamma  \vdash_{\mathcal{M} }  \ottnt{B}   \quad   \phi_{{\mathrm{1}}}  \leq  \phi_{{\mathrm{2}}}  }%
}{
 \phi_{{\mathrm{2}}}  \odot  \Phi  ;  \Gamma  \vdash_{\mathcal{M} }  \ottnt{B} }{%
{\ottdruleNMXXGSubName}{}%
}}

\newcommand{\ottdruleNMXXGWeakName}[0]{\ottdrulename{NM\_GWeak}}
\newcommand{\ottdruleNMXXGWeak}[1]{\ottdrule[#1]{%
\ottpremise{   (  \phi_{{\mathrm{1}}}  \ottsym{,}  \phi_{{\mathrm{2}}}  )   \odot   ( \Phi_{{\mathrm{1}}}  \ottsym{,}  \Phi_{{\mathrm{2}}} )   ;  \Gamma  \vdash_{\mathcal{M} }  \ottnt{B}   \quad   \mathsf{a}  \, \in \, \mathcal{R} }%
}{
  (  \phi_{{\mathrm{1}}}  \ottsym{,}   \mathsf{a}   \ottsym{,}  \phi_{{\mathrm{2}}}  )   \odot   ( \Phi_{{\mathrm{1}}}  \ottsym{,}   \ottnt{X}   \ottsym{,}  \Phi_{{\mathrm{2}}} )   ;  \Gamma  \vdash_{\mathcal{M} }  \ottnt{B} }{%
{\ottdruleNMXXGWeakName}{}%
}}

\newcommand{\ottdruleNMXXGContName}[0]{\ottdrulename{NM\_GCont}}
\newcommand{\ottdruleNMXXGCont}[1]{\ottdrule[#1]{%
\ottpremise{   (  \phi_{{\mathrm{1}}}  \ottsym{,}  \ottmv{r_{{\mathrm{1}}}}  \ottsym{,}  \ottmv{r_{{\mathrm{2}}}}  \ottsym{,}  \phi_{{\mathrm{2}}}  )   \odot   ( \Phi_{{\mathrm{1}}}  \ottsym{,}   \ottnt{X}   \ottsym{,}   \ottnt{X}   \ottsym{,}  \Phi_{{\mathrm{2}}} )   ;  \Gamma  \vdash_{\mathcal{M} }  \ottnt{B}   \quad  \ottsym{(}   \ottmv{r_{{\mathrm{1}}}}  \oplus  \ottmv{r_{{\mathrm{2}}}}   \ottsym{)} \, \in \, \mathcal{R} }%
}{
  (  \phi_{{\mathrm{1}}}  \ottsym{,}   \ottmv{r_{{\mathrm{1}}}}  \oplus  \ottmv{r_{{\mathrm{2}}}}   \ottsym{,}  \phi_{{\mathrm{2}}}  )   \odot   ( \Phi_{{\mathrm{1}}}  \ottsym{,}   \ottnt{X}   \ottsym{,}  \Phi_{{\mathrm{2}}} )   ;  \Gamma  \vdash_{\mathcal{M} }  \ottnt{B} }{%
{\ottdruleNMXXGContName}{}%
}}

\newcommand{\ottdruleNMXXGExName}[0]{\ottdrulename{NM\_GEx}}
\newcommand{\ottdruleNMXXGEx}[1]{\ottdrule[#1]{%
\ottpremise{  (  \phi_{{\mathrm{1}}}  \ottsym{,}  \ottmv{r_{{\mathrm{1}}}}  \ottsym{,}  \ottmv{r_{{\mathrm{2}}}}  \ottsym{,}  \phi_{{\mathrm{2}}}  )   \odot   ( \Phi_{{\mathrm{1}}}  \ottsym{,}   \ottnt{X}   \ottsym{,}   \ottnt{Y}   \ottsym{,}  \Phi_{{\mathrm{2}}} )   ;  \Gamma  \vdash_{\mathcal{M} }  \ottnt{B} }%
}{
  (  \phi_{{\mathrm{1}}}  \ottsym{,}  \ottmv{r_{{\mathrm{2}}}}  \ottsym{,}  \ottmv{r_{{\mathrm{1}}}}  \ottsym{,}  \phi_{{\mathrm{2}}}  )   \odot   ( \Phi_{{\mathrm{1}}}  \ottsym{,}   \ottnt{Y}   \ottsym{,}   \ottnt{X}   \ottsym{,}  \Phi_{{\mathrm{2}}} )   ;  \Gamma  \vdash_{\mathcal{M} }  \ottnt{B} }{%
{\ottdruleNMXXGExName}{}%
}}

\newcommand{\ottdruleNMXXExName}[0]{\ottdrulename{NM\_Ex}}
\newcommand{\ottdruleNMXXEx}[1]{\ottdrule[#1]{%
\ottpremise{ \phi  \odot  \Phi  ;   ( \Gamma_{{\mathrm{1}}}  \ottsym{,}  \ottnt{A}  \ottsym{,}  \ottnt{B}  \ottsym{,}  \Gamma_{{\mathrm{2}}} )   \vdash_{\mathcal{M} }  \ottnt{B} }%
}{
 \phi  \odot  \Phi  ;   ( \Gamma_{{\mathrm{1}}}  \ottsym{,}  \ottnt{B}  \ottsym{,}  \ottnt{A}  \ottsym{,}  \Gamma_{{\mathrm{2}}} )   \vdash_{\mathcal{M} }  \ottnt{B} }{%
{\ottdruleNMXXExName}{}%
}}

\newcommand{\ottdruleTGXXIdName}[0]{\ottdrulename{TG\_Id}}
\newcommand{\ottdruleTGXXId}[1]{\ottdrule[#1]{%
}{
  \mathsf{m}   \odot   \ottmv{x}  :  \ottnt{X}   \vdash_{\mathcal{G} }  \ottmv{x}  :  \ottnt{X} }{%
{\ottdruleTGXXIdName}{}%
}}

\newcommand{\ottdruleTGXXUnitIName}[0]{\ottdrulename{TG\_UnitI}}
\newcommand{\ottdruleTGXXUnitI}[1]{\ottdrule[#1]{%
}{
  \emptyset   \odot   \emptyset   \vdash_{\mathcal{G} }   \mathsf{j}   :   J  }{%
{\ottdruleTGXXUnitIName}{}%
}}

\newcommand{\ottdruleTGXXTenIName}[0]{\ottdrulename{TG\_TenI}}
\newcommand{\ottdruleTGXXTenI}[1]{\ottdrule[#1]{%
\ottpremise{ \phi_{{\mathrm{1}}}  \odot  \Phi_{{\mathrm{1}}}  \vdash_{\mathcal{G} }  \ottnt{t_{{\mathrm{1}}}}  :  \ottnt{X} }%
\ottpremise{ \phi_{{\mathrm{2}}}  \odot  \Phi_{{\mathrm{2}}}  \vdash_{\mathcal{G} }  \ottnt{t_{{\mathrm{2}}}}  :  \ottnt{Y} }%
}{
  (  \phi_{{\mathrm{1}}}  \ottsym{,}  \phi_{{\mathrm{2}}}  )   \odot   ( \Phi_{{\mathrm{1}}}  \ottsym{,}  \Phi_{{\mathrm{2}}} )   \vdash_{\mathcal{G} }  \ottsym{(}  \ottnt{t_{{\mathrm{1}}}}  \ottsym{,}  \ottnt{t_{{\mathrm{2}}}}  \ottsym{)}  :   \ottnt{X}  \boxtimes  \ottnt{Y}  }{%
{\ottdruleTGXXTenIName}{}%
}}

\newcommand{\ottdruleTGXXTenEName}[0]{\ottdrulename{TG\_TenE}}
\newcommand{\ottdruleTGXXTenE}[1]{\ottdrule[#1]{%
\ottpremise{ \phi_{{\mathrm{2}}}  \odot  \Phi_{{\mathrm{2}}}  \vdash_{\mathcal{G} }  \ottnt{t_{{\mathrm{1}}}}  :   \ottnt{X}  \boxtimes  \ottnt{Y}  }%
\ottpremise{   (  \phi_{{\mathrm{1}}}  \ottsym{,}  \ottmv{r}  \ottsym{,}  \ottmv{r}  \ottsym{,}  \phi_{{\mathrm{3}}}  )   \odot   ( \Phi_{{\mathrm{1}}}  \ottsym{,}   \ottnt{X}   \ottsym{,}   \ottnt{Y}   \ottsym{,}  \Phi_{{\mathrm{3}}} )   \vdash_{\mathcal{G} }  \ottnt{t_{{\mathrm{2}}}}  :  \ottnt{Z}   \quad   \vdash  \ottmv{r}  \circledast  \phi_{{\mathrm{2}}}  }%
}{
  (  \phi_{{\mathrm{1}}}  \ottsym{,}    \ottmv{r}  \circledast  \phi_{{\mathrm{2}}}    \ottsym{,}  \phi_{{\mathrm{3}}}  )   \odot   ( \Phi_{{\mathrm{1}}}  \ottsym{,}  \Phi_{{\mathrm{2}}}  \ottsym{,}  \Phi_{{\mathrm{3}}} )   \vdash_{\mathcal{G} }   \mathsf{let}\,( \ottmv{x} , \ottmv{y} ) =  \ottnt{t_{{\mathrm{1}}}} \,\mathsf{in}\, \ottnt{t_{{\mathrm{2}}}}   :  \ottnt{Z} }{%
{\ottdruleTGXXTenEName}{}%
}}

\newcommand{\ottdruleTGXXGrdIName}[0]{\ottdrulename{TG\_GrdI}}
\newcommand{\ottdruleTGXXGrdI}[1]{\ottdrule[#1]{%
\ottpremise{ \phi  \odot  \Phi  ;   \emptyset   \vdash_{\mathcal{M} }  \ottnt{l}  :  \ottnt{B} }%
}{
 \phi  \odot  \Phi  \vdash_{\mathcal{G} }   \mathsf{Grd}\, \ottnt{l}   :   \mathsf{Grd}\, \ottnt{B}  }{%
{\ottdruleTGXXGrdIName}{}%
}}

\newcommand{\ottdruleTGXXSubName}[0]{\ottdrulename{TG\_Sub}}
\newcommand{\ottdruleTGXXSub}[1]{\ottdrule[#1]{%
\ottpremise{  \phi_{{\mathrm{1}}}  \odot  \Phi_{{\mathrm{1}}}  \vdash_{\mathcal{G} }  \ottnt{t}  :  \ottnt{X}   \quad   \phi_{{\mathrm{1}}}  \leq  \phi_{{\mathrm{2}}}  }%
}{
 \phi_{{\mathrm{2}}}  \odot  \Phi_{{\mathrm{2}}}  \vdash_{\mathcal{G} }  \ottnt{t}  :  \ottnt{X} }{%
{\ottdruleTGXXSubName}{}%
}}

\newcommand{\ottdruleTGXXWeakName}[0]{\ottdrulename{TG\_Weak}}
\newcommand{\ottdruleTGXXWeak}[1]{\ottdrule[#1]{%
\ottpremise{   (  \phi_{{\mathrm{1}}}  \ottsym{,}  \phi_{{\mathrm{2}}}  )   \odot   ( \Phi_{{\mathrm{1}}}  \ottsym{,}  \Phi_{{\mathrm{2}}} )   \vdash_{\mathcal{G} }  \ottnt{t}  :  \ottnt{Y}   \quad   \mathsf{a}  \, \in \, \mathcal{R} }%
}{
  (  \phi_{{\mathrm{1}}}  \ottsym{,}   \mathsf{a}   \ottsym{,}  \phi_{{\mathrm{2}}}  )   \odot   ( \Phi_{{\mathrm{1}}}  \ottsym{,}   \ottmv{x}  :  \ottnt{X}   \ottsym{,}  \Phi_{{\mathrm{2}}} )   \vdash_{\mathcal{G} }  \ottnt{t}  :  \ottnt{Y} }{%
{\ottdruleTGXXWeakName}{}%
}}

\newcommand{\ottdruleTGXXContName}[0]{\ottdrulename{TG\_Cont}}
\newcommand{\ottdruleTGXXCont}[1]{\ottdrule[#1]{%
\ottpremise{   (  \phi_{{\mathrm{1}}}  \ottsym{,}  \ottmv{r_{{\mathrm{1}}}}  \ottsym{,}  \ottmv{r_{{\mathrm{2}}}}  \ottsym{,}  \phi_{{\mathrm{2}}}  )   \odot   ( \Phi_{{\mathrm{1}}}  \ottsym{,}   \ottmv{x}  :  \ottnt{X}   \ottsym{,}   \ottmv{y}  :  \ottnt{X}   \ottsym{,}  \Phi_{{\mathrm{2}}} )   \vdash_{\mathcal{G} }  \ottnt{t}  :  \ottnt{Y}   \quad  \ottsym{(}   \ottmv{r_{{\mathrm{1}}}}  \oplus  \ottmv{r_{{\mathrm{2}}}}   \ottsym{)} \, \in \, \mathcal{R} }%
}{
  (  \phi_{{\mathrm{1}}}  \ottsym{,}   \ottmv{r_{{\mathrm{1}}}}  \oplus  \ottmv{r_{{\mathrm{2}}}}   \ottsym{,}  \phi_{{\mathrm{2}}}  )   \odot   ( \Phi_{{\mathrm{1}}}  \ottsym{,}   \ottmv{x}  :  \ottnt{X}   \ottsym{,}  \Phi_{{\mathrm{2}}} )   \vdash_{\mathcal{G} }  \ottsym{[}  \ottmv{x}  \ottsym{/}  \ottmv{y}  \ottsym{]}  \ottnt{t}  :  \ottnt{Y} }{%
{\ottdruleTGXXContName}{}%
}}

\newcommand{\ottdruleTGXXExName}[0]{\ottdrulename{TG\_Ex}}
\newcommand{\ottdruleTGXXEx}[1]{\ottdrule[#1]{%
\ottpremise{  (  \phi_{{\mathrm{1}}}  \ottsym{,}  \ottmv{r_{{\mathrm{1}}}}  \ottsym{,}  \ottmv{r_{{\mathrm{2}}}}  \ottsym{,}  \phi_{{\mathrm{2}}}  )   \odot   ( \Phi_{{\mathrm{1}}}  \ottsym{,}   \ottmv{x}  :  \ottnt{X}   \ottsym{,}   \ottmv{y}  :  \ottnt{Y}   \ottsym{,}  \Phi_{{\mathrm{2}}} )   \vdash_{\mathcal{G} }  \ottnt{t}  :  \ottnt{Z} }%
}{
  (  \phi_{{\mathrm{1}}}  \ottsym{,}  \ottmv{r_{{\mathrm{2}}}}  \ottsym{,}  \ottmv{r_{{\mathrm{1}}}}  \ottsym{,}  \phi_{{\mathrm{2}}}  )   \odot   ( \Phi_{{\mathrm{1}}}  \ottsym{,}   \ottmv{y}  :  \ottnt{Y}   \ottsym{,}   \ottmv{x}  :  \ottnt{X}   \ottsym{,}  \Phi_{{\mathrm{2}}} )   \vdash_{\mathcal{G} }  \ottnt{t}  :  \ottnt{Z} }{%
{\ottdruleTGXXExName}{}%
}}

\newcommand{\ottdruleTMXXIdName}[0]{\ottdrulename{TM\_Id}}
\newcommand{\ottdruleTMXXId}[1]{\ottdrule[#1]{%
}{
  \emptyset   \odot   \emptyset   ;  \ottmv{x}  \ottsym{:}  \ottnt{A}  \vdash_{\mathcal{M} }  \ottmv{x}  :  \ottnt{A} }{%
{\ottdruleTMXXIdName}{}%
}}

\newcommand{\ottdruleTMXXUnitIName}[0]{\ottdrulename{TM\_UnitI}}
\newcommand{\ottdruleTMXXUnitI}[1]{\ottdrule[#1]{%
}{
  \emptyset   \odot   \emptyset   ;   \emptyset   \vdash_{\mathcal{M} }   \mathsf{i}   :   I  }{%
{\ottdruleTMXXUnitIName}{}%
}}

\newcommand{\ottdruleTMXXTenIName}[0]{\ottdrulename{TM\_TenI}}
\newcommand{\ottdruleTMXXTenI}[1]{\ottdrule[#1]{%
\ottpremise{ \phi_{{\mathrm{1}}}  \odot  \Phi_{{\mathrm{1}}}  ;  \Gamma_{{\mathrm{1}}}  \vdash_{\mathcal{M} }  \ottnt{l_{{\mathrm{1}}}}  :  \ottnt{A} }%
\ottpremise{ \phi_{{\mathrm{2}}}  \odot  \Phi_{{\mathrm{2}}}  ;  \Gamma_{{\mathrm{2}}}  \vdash_{\mathcal{M} }  \ottnt{l_{{\mathrm{2}}}}  :  \ottnt{B} }%
}{
  (  \phi_{{\mathrm{1}}}  \ottsym{,}  \phi_{{\mathrm{2}}}  )   \odot   ( \Phi_{{\mathrm{1}}}  \ottsym{,}  \Phi_{{\mathrm{2}}} )   ;   ( \Gamma_{{\mathrm{1}}}  \ottsym{,}  \Gamma_{{\mathrm{2}}} )   \vdash_{\mathcal{M} }  \ottsym{(}  \ottnt{l_{{\mathrm{1}}}}  \ottsym{,}  \ottnt{l_{{\mathrm{2}}}}  \ottsym{)}  :   \ottnt{A}  \otimes  \ottnt{B}  }{%
{\ottdruleTMXXTenIName}{}%
}}

\newcommand{\ottdruleTMXXTenEName}[0]{\ottdrulename{TM\_TenE}}
\newcommand{\ottdruleTMXXTenE}[1]{\ottdrule[#1]{%
\ottpremise{ \phi_{{\mathrm{2}}}  \odot  \Phi_{{\mathrm{2}}}  ;  \Gamma_{{\mathrm{2}}}  \vdash_{\mathcal{M} }  \ottnt{l_{{\mathrm{1}}}}  :   \ottnt{A}  \otimes  \ottnt{B}  }%
\ottpremise{ \phi_{{\mathrm{1}}}  \odot  \Phi_{{\mathrm{1}}}  ;  \Gamma_{{\mathrm{1}}}  \ottsym{,}  \ottnt{A}  \ottsym{,}  \ottnt{B}  \ottsym{,}  \Gamma_{{\mathrm{3}}}  \vdash_{\mathcal{M} }  \ottnt{l_{{\mathrm{2}}}}  :  \ottnt{C} }%
}{
  (  \phi_{{\mathrm{1}}}  \ottsym{,}  \phi_{{\mathrm{2}}}  )   \odot   ( \Phi_{{\mathrm{1}}}  \ottsym{,}  \Phi_{{\mathrm{2}}} )   ;   ( \Gamma_{{\mathrm{1}}}  \ottsym{,}  \Gamma_{{\mathrm{2}}}  \ottsym{,}  \Gamma_{{\mathrm{3}}} )   \vdash_{\mathcal{M} }   \mathsf{let}\,( \ottmv{x} , \ottmv{y} ) =  \ottnt{l_{{\mathrm{1}}}} \,\mathsf{in}\, \ottnt{l_{{\mathrm{2}}}}   :  \ottnt{C} }{%
{\ottdruleTMXXTenEName}{}%
}}

\newcommand{\ottdruleTMXXImpIName}[0]{\ottdrulename{TM\_ImpI}}
\newcommand{\ottdruleTMXXImpI}[1]{\ottdrule[#1]{%
\ottpremise{ \phi  \odot  \Phi  ;   ( \Gamma  \ottsym{,}  \ottnt{A} )   \vdash_{\mathcal{M} }  \ottnt{l}  :  \ottnt{B} }%
}{
 \phi  \odot  \Phi  ;  \Gamma  \vdash_{\mathcal{M} }   \lambda  \ottmv{x} . \ottnt{l}   :   \ottnt{A}  \multimap  \ottnt{B}  }{%
{\ottdruleTMXXImpIName}{}%
}}

\newcommand{\ottdruleTMXXImpEName}[0]{\ottdrulename{TM\_ImpE}}
\newcommand{\ottdruleTMXXImpE}[1]{\ottdrule[#1]{%
\ottpremise{ \phi_{{\mathrm{2}}}  \odot  \Phi_{{\mathrm{2}}}  ;  \Gamma_{{\mathrm{2}}}  \vdash_{\mathcal{M} }  \ottnt{l_{{\mathrm{2}}}}  :  \ottnt{A} }%
\ottpremise{ \phi_{{\mathrm{1}}}  \odot  \Phi_{{\mathrm{1}}}  ;  \Gamma_{{\mathrm{1}}}  \vdash_{\mathcal{M} }  \ottnt{l_{{\mathrm{1}}}}  :   \ottnt{A}  \multimap  \ottnt{B}  }%
}{
  (  \phi_{{\mathrm{1}}}  \ottsym{,}  \phi_{{\mathrm{2}}}  )   \odot   ( \Phi_{{\mathrm{1}}}  \ottsym{,}  \Phi_{{\mathrm{2}}} )   ;   ( \Gamma_{{\mathrm{1}}}  \ottsym{,}  \Gamma_{{\mathrm{2}}} )   \vdash_{\mathcal{M} }   \ottnt{l_{{\mathrm{1}}}} \, \ottnt{l_{{\mathrm{2}}}}   :  \ottnt{B} }{%
{\ottdruleTMXXImpEName}{}%
}}

\newcommand{\ottdruleTMXXLinIName}[0]{\ottdrulename{TM\_LinI}}
\newcommand{\ottdruleTMXXLinI}[1]{\ottdrule[#1]{%
\ottpremise{  \phi  \odot  \Phi  \vdash_{\mathcal{G} }  \ottnt{t}  :  \ottnt{X}   \quad   \vdash  \ottmv{r}  \circledast  \phi  }%
}{
  \ottmv{r}  \circledast  \phi   \odot  \Phi  ;   \emptyset   \vdash_{\mathcal{M} }   \mathsf{Lin}\, \ottnt{t}   :   \mathsf{Lin}_{ \ottmv{r} }\, \ottnt{X}  }{%
{\ottdruleTMXXLinIName}{}%
}}

\newcommand{\ottdruleTMXXLinEName}[0]{\ottdrulename{TM\_LinE}}
\newcommand{\ottdruleTMXXLinE}[1]{\ottdrule[#1]{%
\ottpremise{ \phi_{{\mathrm{2}}}  \odot  \Phi_{{\mathrm{2}}}  ;  \Gamma_{{\mathrm{2}}}  \vdash_{\mathcal{M} }  \ottnt{l_{{\mathrm{1}}}}  :   \mathsf{Lin}_{ \ottmv{r} }\, \ottnt{X}  }%
\ottpremise{  (  \phi_{{\mathrm{1}}}  \ottsym{,}  \ottmv{r}  \ottsym{,}  \phi_{{\mathrm{2}}}  )   \odot   ( \Phi_{{\mathrm{1}}}  \ottsym{,}   \ottnt{X}   \ottsym{,}  \Phi_{{\mathrm{3}}} )   ;  \Gamma_{{\mathrm{1}}}  \vdash_{\mathcal{M} }  \ottnt{l_{{\mathrm{2}}}}  :  \ottnt{B} }%
}{
  (  \phi_{{\mathrm{1}}}  \ottsym{,}  \phi_{{\mathrm{2}}}  \ottsym{,}  \phi_{{\mathrm{3}}}  )   \odot   ( \Phi_{{\mathrm{1}}}  \ottsym{,}  \Phi_{{\mathrm{2}}}  \ottsym{,}  \Phi_{{\mathrm{3}}} )   ;   ( \Gamma_{{\mathrm{1}}}  \ottsym{,}  \Gamma_{{\mathrm{2}}} )   \vdash_{\mathcal{M} }   \mathsf{let}\,\mathsf{Lin}\, \ottmv{x}  =  \ottnt{l_{{\mathrm{1}}}} \,\mathsf{in}\, \ottnt{l_{{\mathrm{2}}}}   :  \ottnt{B} }{%
{\ottdruleTMXXLinEName}{}%
}}

\newcommand{\ottdruleTMXXGrdEName}[0]{\ottdrulename{TM\_GrdE}}
\newcommand{\ottdruleTMXXGrdE}[1]{\ottdrule[#1]{%
\ottpremise{ \phi  \odot  \Phi  \vdash_{\mathcal{G} }  \ottnt{t}  :   \mathsf{Grd}\, \ottnt{A}  }%
}{
 \phi  \odot  \Phi  ;   \emptyset   \vdash_{\mathcal{M} }   \mathsf{Ungrd}\, \ottnt{t}   :  \ottnt{A} }{%
{\ottdruleTMXXGrdEName}{}%
}}

\newcommand{\ottdruleTMXXGSubName}[0]{\ottdrulename{TM\_GSub}}
\newcommand{\ottdruleTMXXGSub}[1]{\ottdrule[#1]{%
\ottpremise{  \phi_{{\mathrm{1}}}  \odot  \Phi  ;  \Gamma  \vdash_{\mathcal{M} }  \ottnt{l}  :  \ottnt{B}   \quad   \phi_{{\mathrm{1}}}  \leq  \phi_{{\mathrm{2}}}  }%
}{
 \phi_{{\mathrm{2}}}  \odot  \Phi  ;  \Gamma  \vdash_{\mathcal{M} }  \ottnt{l}  :  \ottnt{B} }{%
{\ottdruleTMXXGSubName}{}%
}}

\newcommand{\ottdruleTMXXGWeakName}[0]{\ottdrulename{TM\_GWeak}}
\newcommand{\ottdruleTMXXGWeak}[1]{\ottdrule[#1]{%
\ottpremise{   (  \phi_{{\mathrm{1}}}  \ottsym{,}  \phi_{{\mathrm{2}}}  )   \odot   ( \Phi_{{\mathrm{1}}}  \ottsym{,}  \Phi_{{\mathrm{2}}} )   ;  \Gamma  \vdash_{\mathcal{M} }  \ottnt{l}  :  \ottnt{B}   \quad   \mathsf{a}  \, \in \, \mathcal{R} }%
}{
  (  \phi_{{\mathrm{1}}}  \ottsym{,}   \mathsf{a}   \ottsym{,}  \phi_{{\mathrm{2}}}  )   \odot   ( \Phi_{{\mathrm{1}}}  \ottsym{,}   \ottnt{X}   \ottsym{,}  \Phi_{{\mathrm{2}}} )   ;  \Gamma  \vdash_{\mathcal{M} }  \ottnt{l}  :  \ottnt{B} }{%
{\ottdruleTMXXGWeakName}{}%
}}

\newcommand{\ottdruleTMXXGContName}[0]{\ottdrulename{TM\_GCont}}
\newcommand{\ottdruleTMXXGCont}[1]{\ottdrule[#1]{%
\ottpremise{   (  \phi_{{\mathrm{1}}}  \ottsym{,}  \ottmv{r_{{\mathrm{1}}}}  \ottsym{,}  \ottmv{r_{{\mathrm{2}}}}  \ottsym{,}  \phi_{{\mathrm{2}}}  )   \odot   ( \Phi_{{\mathrm{1}}}  \ottsym{,}   \ottnt{X}   \ottsym{,}   \ottnt{X}   \ottsym{,}  \Phi_{{\mathrm{2}}} )   ;  \Gamma  \vdash_{\mathcal{M} }  \ottnt{l}  :  \ottnt{B}   \quad  \ottsym{(}   \ottmv{r_{{\mathrm{1}}}}  \oplus  \ottmv{r_{{\mathrm{2}}}}   \ottsym{)} \, \in \, \mathcal{R} }%
}{
  (  \phi_{{\mathrm{1}}}  \ottsym{,}   \ottmv{r_{{\mathrm{1}}}}  \oplus  \ottmv{r_{{\mathrm{2}}}}   \ottsym{,}  \phi_{{\mathrm{2}}}  )   \odot   ( \Phi_{{\mathrm{1}}}  \ottsym{,}   \ottnt{X}   \ottsym{,}  \Phi_{{\mathrm{2}}} )   ;  \Gamma  \vdash_{\mathcal{M} }  \ottsym{[}  \ottmv{x}  \ottsym{/}  \ottmv{y}  \ottsym{]}  \ottnt{l}  :  \ottnt{B} }{%
{\ottdruleTMXXGContName}{}%
}}

\newcommand{\ottdruleTMXXGExName}[0]{\ottdrulename{TM\_GEx}}
\newcommand{\ottdruleTMXXGEx}[1]{\ottdrule[#1]{%
\ottpremise{  (  \phi_{{\mathrm{1}}}  \ottsym{,}  \ottmv{r_{{\mathrm{1}}}}  \ottsym{,}  \ottmv{r_{{\mathrm{2}}}}  \ottsym{,}  \phi_{{\mathrm{2}}}  )   \odot   ( \Phi_{{\mathrm{1}}}  \ottsym{,}   \ottnt{X}   \ottsym{,}   \ottnt{Y}   \ottsym{,}  \Phi_{{\mathrm{2}}} )   ;  \Gamma  \vdash_{\mathcal{M} }  \ottnt{l}  :  \ottnt{B} }%
}{
  (  \phi_{{\mathrm{1}}}  \ottsym{,}  \ottmv{r_{{\mathrm{2}}}}  \ottsym{,}  \ottmv{r_{{\mathrm{1}}}}  \ottsym{,}  \phi_{{\mathrm{2}}}  )   \odot   ( \Phi_{{\mathrm{1}}}  \ottsym{,}   \ottnt{Y}   \ottsym{,}   \ottnt{X}   \ottsym{,}  \Phi_{{\mathrm{2}}} )   ;  \Gamma  \vdash_{\mathcal{M} }  \ottnt{l}  :  \ottnt{B} }{%
{\ottdruleTMXXGExName}{}%
}}

\newcommand{\ottdruleTMXXExName}[0]{\ottdrulename{TM\_Ex}}
\newcommand{\ottdruleTMXXEx}[1]{\ottdrule[#1]{%
\ottpremise{ \phi  \odot  \Phi  ;   ( \Gamma_{{\mathrm{1}}}  \ottsym{,}  \ottnt{A}  \ottsym{,}  \ottnt{B}  \ottsym{,}  \Gamma_{{\mathrm{2}}} )   \vdash_{\mathcal{M} }  \ottnt{l}  :  \ottnt{B} }%
}{
 \phi  \odot  \Phi  ;   ( \Gamma_{{\mathrm{1}}}  \ottsym{,}  \ottnt{B}  \ottsym{,}  \ottnt{A}  \ottsym{,}  \Gamma_{{\mathrm{2}}} )   \vdash_{\mathcal{M} }  \ottnt{l}  :  \ottnt{B} }{%
{\ottdruleTMXXExName}{}%
}}

\renewcommand{\ottdrule}[4][]{{\displaystyle\frac{\begin{array}{l}#2\end{array}}{#3}\,{\footnotesize #4}}}

\renewcommand{\ottdruleVMXXXEmptyName}{\textbf{empty}}
\renewcommand{\ottdruleVMXXXExtName}{\textbf{ext}}

\renewcommand{\ottdruleSGXXidName}{\textbf{id}}
\renewcommand{\ottdruleSGXXUnitRName}{\textbf{unit}_R}
\renewcommand{\ottdruleSGXXTenLName}{\boxtimes_L}
\renewcommand{\ottdruleSGXXTenRName}{\boxtimes_R}
\renewcommand{\ottdruleSGXXGrdRName}{\mathsf{Grd}_R}
\renewcommand{\ottdruleSGXXCutName}{\textbf{cut}}
\renewcommand{\ottdruleSGXXSubName}{\textbf{sub}}
\renewcommand{\ottdruleSGXXWeakName}{\textbf{weak}}
\renewcommand{\ottdruleSGXXContName}{\textbf{contr}}
\renewcommand{\ottdruleSGXXExName}{\textbf{ex}}

\renewcommand{\ottdruleSMXXIdName}{\textbf{id}}
\renewcommand{\ottdruleSMXXUnitRName}{\textbf{unit}_R}
\renewcommand{\ottdruleSMXXGTenLName}{\boxtimes_L}
\renewcommand{\ottdruleSMXXTenLName}{\otimes_L}
\renewcommand{\ottdruleSMXXTenRName}{\otimes_R}
\renewcommand{\ottdruleSMXXImpLName}{\multimap_L}
\renewcommand{\ottdruleSMXXImpRName}{\multimap_R}
\renewcommand{\ottdruleSMXXGrdLName}{\mathsf{Grd}_L}
\renewcommand{\ottdruleSMXXLinLName}{\mathsf{Lin}_L}
\renewcommand{\ottdruleSMXXLinRName}{\mathsf{Lin}_R}
\renewcommand{\ottdruleSMXXCutName}{\textbf{cut}}
\renewcommand{\ottdruleSMXXGCutName}{\textbf{cut}_G}
\renewcommand{\ottdruleSMXXGSubName}{\textbf{sub}}
\renewcommand{\ottdruleSMXXGWeakName}{\textbf{weak}}
\renewcommand{\ottdruleSMXXGContName}{\textbf{contr}}
\renewcommand{\ottdruleSMXXGExName}{\textbf{ex}_G}
\renewcommand{\ottdruleSMXXExName}{\textbf{ex}}

\renewcommand{\ottdruleNGXXIdName}{\textbf{id}}
\renewcommand{\ottdruleNGXXUnitIName}{\textbf{unit}_I}
\renewcommand{\ottdruleNGXXTenEName}{\boxtimes_E}
\renewcommand{\ottdruleNGXXTenIName}{\boxtimes_I}
\renewcommand{\ottdruleNGXXGrdIName}{\mathsf{Grd}_I}
\renewcommand{\ottdruleNGXXSubName}{\textbf{sub}}
\renewcommand{\ottdruleNGXXWeakName}{\textbf{weak}}
\renewcommand{\ottdruleNGXXContName}{\textbf{contr}}
\renewcommand{\ottdruleNGXXExName}{\textbf{ex}}

\renewcommand{\ottdruleNMXXIdName}{\textbf{id}}
\renewcommand{\ottdruleNMXXUnitIName}{\textbf{unit}_I}
\renewcommand{\ottdruleNMXXTenEName}{\otimes_E}
\renewcommand{\ottdruleNMXXTenIName}{\otimes_I}
\renewcommand{\ottdruleNMXXImpEName}{\multimap_E}
\renewcommand{\ottdruleNMXXImpIName}{\multimap_I}
\renewcommand{\ottdruleNMXXGrdEName}{\mathsf{Grd}_E}
\renewcommand{\ottdruleNMXXLinEName}{\mathsf{Lin}_E}
\renewcommand{\ottdruleNMXXLinIName}{\mathsf{Lin}_I}
\renewcommand{\ottdruleNMXXGSubName}{\textbf{sub}}
\renewcommand{\ottdruleNMXXGWeakName}{\textbf{weak}}
\renewcommand{\ottdruleNMXXGContName}{\textbf{contr}}
\renewcommand{\ottdruleNMXXGExName}{\textbf{ex}_G}
\renewcommand{\ottdruleNMXXExName}{\textbf{ex}}

\renewcommand{\ottdruleTGXXIdName}{\textbf{id}}
\renewcommand{\ottdruleTGXXUnitIName}{\textbf{unit}_I}
\renewcommand{\ottdruleTGXXTenEName}{\boxtimes_E}
\renewcommand{\ottdruleTGXXTenIName}{\boxtimes_I}
\renewcommand{\ottdruleTGXXGrdIName}{\mathsf{Grd}_I}
\renewcommand{\ottdruleTGXXSubName}{\textbf{sub}}
\renewcommand{\ottdruleTGXXWeakName}{\textbf{weak}}
\renewcommand{\ottdruleTGXXContName}{\textbf{contr}}
\renewcommand{\ottdruleTGXXExName}{\textbf{ex}}

\renewcommand{\ottdruleTMXXIdName}{\textbf{id}}
\renewcommand{\ottdruleTMXXUnitIName}{\textbf{unit}_I}
\renewcommand{\ottdruleTMXXTenEName}{\otimes_E}
\renewcommand{\ottdruleTMXXTenIName}{\otimes_I}
\renewcommand{\ottdruleTMXXImpEName}{\multimap_E}
\renewcommand{\ottdruleTMXXImpIName}{\multimap_I}
\renewcommand{\ottdruleTMXXGrdEName}{\mathsf{Grd}_E}
\renewcommand{\ottdruleTMXXLinEName}{\mathsf{Lin}_E}
\renewcommand{\ottdruleTMXXLinIName}{\mathsf{Lin}_I}
\renewcommand{\ottdruleTMXXGSubName}{\textbf{sub}}
\renewcommand{\ottdruleTMXXGWeakName}{\textbf{weak}}
\renewcommand{\ottdruleTMXXGContName}{\textbf{contr}}
\renewcommand{\ottdruleTMXXGExName}{\textbf{ex}_G}
\renewcommand{\ottdruleTMXXExName}{\textbf{ex}}

\title{Grading Adjoint Logic}
\author{Harley Eades III
\institute{School of Computer and Cyber Sciences\\
Augusta University}
\email{harley.eades@gmail.com}
\and
Dominic Orchard
\institute{School of Computing\\
University of Kent}
\email{d.a.orchard@kent.ac.uk}
}
\def\authorrunning{H. Eades III \& D. Orchard}

\def\titlerunning{Graded Adjoint Logic}

\begin{document}

\maketitle

Girard's linear logic~\cite{Girard:1987} has lead to many applications
in logic, mathematics, and computer science.  Recently, linear logic
has seen two refinements: Adjoint Logic and Graded Modal Logic.

\textbf{Adjoint Logic.} Adjoint logic is a generalization of Benton's
beautiful Linear/Non-linear (LNL) logic.  This consists of two
fragments: intuitionistic non-linear logic $ \Phi  \vdash_{\cat{I} }  \ottnt{X} $ and a mixed
fragment of intuitionistic linear logic with non-linear hypotheses
$ \Phi ; \Gamma  \vdash_{\cat{L} }  \ottnt{A} $. These two fragments are connected by a pair of
modalities $ \mathsf{Lin} (  \ottnt{X}  ) $ and $ \mathsf{Mny}( \ottnt{A} ) $ which form an adjunction.  The
former takes a non-linear formula, $\ottnt{X}$, and brings it into the
linear fragment, while $ \mathsf{Mny}( \ottnt{A} ) $ brings a linear formula, $\ottnt{A}$,
into the non-linear fragment.  Girard's
of-course modality can be recovered by $!A =  \mathsf{Lin} (   \mathsf{Mny}( \ottnt{A} )   ) $.  Breaking
the of-course modality into two modalities and allowing linear
logic to be mixed with non-linear logic has been very fruitful, and so
a natural question is ``is it possible to build LNL-like logics for
other substructural logics?''

The non-linear fragment can be viewed as linear logic with the
addition of structural rules for weakening and contraction.  If we
remove one of these rules, then we obtain a different substructural system.
Pruiksma et al.~\cite{Pruiksma:2018} proposed a flexible approach
in the form of a new logic called Adjoint logic.  This system
restructures LNL logic so that formulas can be annotated
with a \emph{mode} $m$ that indicates via a labelling $\sigma$ which structural rules are allowed
for that formula.  For example, if we take a mode $m$ whose only
structural rule is weakening, denoted $\sigma(\mathsf{m}) = \{\mathsf{W}\}$, then a
formula $A_{\mathsf{m}}$ is an affine formula: it can be used
zero or one times.  The logic is then parameterized by a theory of
modes.  Different instantiations of the mode structure yield
different kinds of adjoint logic.  Note that when we refer to Adjoint
Logic we are referring to the work of Pruiksma et
al.~\cite{Pruiksma:2018} and not the more general work of Licata et
al.~\cite{Licata:2019}.

\textbf{Graded Modal Logic.} In contrast, Graded Modal
Logics~\cite{Brunel:2014,Gaboardi:2016,Katsumata:2018,Orchard:2019}
refine linear logic by replacing the of-course modality, $!A$, with a
\emph{graded necessity modality}, $ \Box_{ \ottmv{r} } \ottnt{A} $, which annotates
formulas with a usage constraint, $\ottmv{r}$, called a \emph{grade}
drawn from a semiring $(\cat{R}, \ottmv{m}, \oast,  \mathsf{a} ,
\oplus)$, parameterizing the logic.  The multiplicative
structure of the semiring is used for composition of proofs, and
the additive structure is used to control the usage of the graded
structural rules. Typically, hypotheses are annotated with a grade.
We write $\gamma \odot \Gamma \vdash A$ where $\gamma$ is a context of
grades whose structure matches $\Gamma$. Structural rules
are then:
\[ \small
\begin{array}{cc}
  \inferrule* [flushleft,right={\footnotesize \textbf{weak}}] {
  (\gamma_1,\gamma_2) \odot (\Gamma_{{\mathrm{1}}},\Gamma_{{\mathrm{2}}}) \vdash \ottnt{X_{{\mathrm{2}}}}
}{(\gamma_1, \mathsf{a} ,\gamma_2) \odot (\Gamma_{{\mathrm{1}}},\ottnt{X_{{\mathrm{1}}}},\Gamma_{{\mathrm{2}}}) \vdash \ottnt{X_{{\mathrm{2}}}}}
&
\inferrule* [flushleft,right={\footnotesize \textbf{contr}}] {
  (\gamma_1,r,s,\gamma_2) \odot (\Gamma_{{\mathrm{1}}},\ottnt{X_{{\mathrm{1}}}},\ottnt{X_{{\mathrm{1}}}},\Gamma_{{\mathrm{2}}}) \vdash \ottnt{X_{{\mathrm{2}}}}
}{(\gamma_1, \ottmv{r}  \oplus  \ottmv{s} ,\gamma_2) \odot (\Gamma_{{\mathrm{1}}},\ottnt{X_{{\mathrm{1}}}},\Gamma_{{\mathrm{2}}}) \vdash \ottnt{X_{{\mathrm{2}}}}}
\end{array}
\]
One can view graded modal logics as providing a means of
quantitatively controlling the use of the structural rules.  For
example, for $(\mathbb{N},1,*,0,+)$ as the semiring, a
graded formula $ \Box_{ \ottmv{r} } \ottnt{A} $ for $\ottmv{r} \in \mathbb{N}$ can be used
$\ottmv{r}$-times in a proof.  If we take the semiring to be
$(\{\infty\}, \infty, (\lambda \ottmv{r_{{\mathrm{1}}}}.\lambda \ottmv{r_{{\mathrm{2}}}}.\infty),
              \infty, (\lambda \ottmv{r_{{\mathrm{1}}}}.\lambda \ottmv{r_{{\mathrm{2}}}}.\infty))$, then the logic degenerates to
non-linear logic. A pre-ordering on $\cat{R}$ can be included yielding
further control.

\textbf{Our Contribution.}  Adjoint logic adds modes to \emph{control
  which} structural rules are allowed, and Graded Modal Logic adds
grades to \emph{control how} the structural rules are used.
An open question is whether these two perspectives can be brought
together under one roof.
 We propose \emph{Graded Adjoint Logic} a graded
modal logic in the style of Benton's LNL logic, but where the semiring
structure has been generalized to support isolating structural rules
to particular modes.

\noindent
The key idea is to generalize semirings to \emph{pointed
  semirings} allowing the semiring structure to be partial.

\begin{definition}
  \label{def:pointed-monoid}
  A \textbf{pointed monoid} is a monoid in $\PSet$. That is, a
  pointed monoid $(M_*,\mathsf{e}_*,\boxtimes_*)$ comprises a
  pointed set $M_* = M \cup \{*\}$, an identity element $\mathsf{e}_* : I_* \mto
  M_*$, and a multiplication $\boxtimes_* : M_* \otimes_* M_* \mto M_*$
  subject to partial associativity and identity axioms (e.g.,
for $\# \in I_*$,
if $\mathsf{e}_*(\#) \in M$ and
$(\mathsf{e}_* \boxtimes_* s) \in M$ then $\mathsf{e}_* \boxtimes_* s
= s$).
%
A pointed monoid is \textbf{unital} iff $\mathsf{e}_*(\#) \in
M$\footnote{Unital pointed monoids are also known as partial monoids
  and pointed semirings are also known as partial semirings in the
  literature.}.
\end{definition}
\noindent
\begin{definition}
  \label{def:resource-algebra}
  A \textbf{pointed semiring} $(\mathcal{R}, \ottmv{m}, \oast,  \mathsf{a} , \oplus)$
  comprises a set $\cat{R}$,
  a pointed unital monoid $(\cat{R}_*, \ottmv{m}, \oast)$, and
  a pointed commutative monoid $(\cat{R}_*,  \mathsf{a} , \oplus)$,
  with partial absorption and distributivity axioms.
\end{definition}
\noindent
In our system, the graded structural rules are then generalized to the following:
\[ \small
\begin{array}{cc}
  \inferrule* [flushleft,right={\footnotesize \textbf{weak}}] {
    (\gamma_1,\gamma_2) \odot (\Gamma_{{\mathrm{1}}},\Gamma_{{\mathrm{2}}}) \vdash \ottnt{X_{{\mathrm{2}}}}\\
     \mathsf{a}  \, \in \, \mathcal{R}
}{(\gamma_1, \mathsf{a} ,\gamma_2) \odot (\Gamma_{{\mathrm{1}}},\ottnt{X_{{\mathrm{1}}}},\Gamma_{{\mathrm{2}}}) \vdash \ottnt{X_{{\mathrm{2}}}}}
&
\inferrule* [flushleft,right={\footnotesize \textbf{contr}}] {
  (\gamma_1,r,s,\gamma_2) \odot (\Gamma_{{\mathrm{1}}},\ottnt{X_{{\mathrm{1}}}},\ottnt{X_{{\mathrm{1}}}},\Gamma_{{\mathrm{2}}}) \vdash \ottnt{X_{{\mathrm{2}}}}\\
  \ottsym{(}   \ottmv{r}  \oplus  \ottmv{s}   \ottsym{)} \, \in \, \mathcal{R}
}{(\gamma_1, \ottmv{r}  \oplus  \ottmv{s} ,\gamma_2) \odot (\Gamma_{{\mathrm{1}}},\ottnt{X_{{\mathrm{1}}}},\Gamma_{{\mathrm{2}}}) \vdash \ottnt{X_{{\mathrm{2}}}}}
\end{array}
\]
The partiality of the structure of the semiring makes it possible for
the elements of $\mathcal{R}$ to double as both grades and modes.  For
example, the pointed semiring $(\{ \mathsf{l}, \mathsf{w}, \mathsf{c}
\}, \mathsf{l}, \oast, \mathsf{w}, \oplus)$ for the
combination of linear, affine, and relevance logic is defined as
follows:
\begin{center}
  \small
\begin{math}
  \begin{array}{lll}
    \begin{array}{|c|c|c|c|c|c|c|c|c|c|c|c|c|c|c|c|}
    \hline
    \ottmv{r_{{\mathrm{1}}}}      & \mathsf{l} & \mathsf{l} & \mathsf{w} & \mathsf{c} & \mathsf{l} & \mathsf{w} & \mathsf{c} & \mathsf{w} & \mathsf{c}\\
    \hline
    \ottmv{r_{{\mathrm{2}}}}      & \mathsf{w} & \mathsf{c} & \mathsf{l} & \mathsf{l} & \mathsf{l} & \mathsf{w} & \mathsf{c} & \mathsf{c} & \mathsf{w}\\
    \hline
     \ottmv{r_{{\mathrm{1}}}}  \circledast  \ottmv{r_{{\mathrm{2}}}}  & \mathsf{w} & \mathsf{c} & \mathsf{w} & \mathsf{c} & \mathsf{l} & \mathsf{w} & \mathsf{c} & * & *\\
    \hline
    \end{array}
    &
    \begin{array}{|c|c|c|c|c|c|c|c|c|c|c|c|c|c|c|c|}
    \hline
    \ottmv{r_{{\mathrm{1}}}}      & \mathsf{l} & \mathsf{l} & \mathsf{w} & \mathsf{c} & \mathsf{l} & \mathsf{w} & \mathsf{c} & \mathsf{w} & \mathsf{c}\\
    \hline
    \ottmv{r_{{\mathrm{2}}}}      & \mathsf{w} & \mathsf{c} & \mathsf{l} & \mathsf{l} & \mathsf{l} & \mathsf{w} & \mathsf{c} & \mathsf{c} & \mathsf{w}\\
    \hline
     \ottmv{r_{{\mathrm{1}}}}  \oplus  \ottmv{r_{{\mathrm{2}}}}  & * & * & * & * & * & * & \mathsf{c} & * & *\\
    \hline
  \end{array}
  \end{array}
\end{math}
\end{center}
Here we designate the grades $\mathsf{l}$, $\mathsf{w}$, and
$\mathsf{c}$ as three modes.  The mode $\mathsf{l}$ stands for
``linear'' and allows no structural rules, the mode $\mathsf{w}$
allows weakening (because it is the additive identity), but not
contraction, however the mode $\mathsf{c}$ allows contraction, but not
weakening. Thus, addition is only defined for the mode $\mathsf{c}$
since we separate contraction from weakening (denoted by the partial
additive identity $\mathsf{w}$). Other more fine-grained combinations
are possible, e.g., taking $\mathcal{P}(\{\mathsf{l}, \mathsf{w},
\mathsf{c}\})$ as the underlying set.

We have developed a sequent calculus, natural deduction, and term
assignment for Graded Adjoint Logic. There definitions are all
summarized in Appendix~\ref{sec:full_systems}.  Each of these systems
consist of two fragments: the graded fragment $ \phi  \odot  \Phi  \vdash_{\mathcal{G} }  \ottnt{X} $ and
the mixed graded/linear fragment $ \phi  \odot  \Phi  ;  \Gamma  \vdash_{\mathcal{M} }  \ottnt{A} $.  Then these two
fragments are connected via adjoint modalities:
\[\small
\begin{array}{llll}
  \ottdruleSGXXGrdR{} & \ottdruleSMXXGrdL{} \\[13px]
  \ottdruleSMXXLinL{} & \ottdruleSMXXLinR{}
\end{array}
\]
Note that $\mathsf{Lin}$ is now an indexed family of modalities $ \mathsf{Lin}_{ \ottmv{r} }\, \ottnt{X} $.  The side condition $ \vdash  \ottmv{r}  \circledast  \phi $ in the rightmost
rule ensures that the scalar multiplication (of a vector of grades
$\phi$) is defined.  Using these rules we can now define a graded
modality by $ \Box_{ \ottmv{r} } \ottnt{A}  =  \mathsf{Lin}_{ \ottmv{r} }\, \ottsym{(}   \mathsf{Grd}\, \ottnt{A}   \ottsym{)} $.  Graded Adjoint Logic
is more general than Graded Modal Logic, because every semiring is a
pointed semiring where all operations are defined.  In addition, this
system is more general than Adjoint Logic, because the proposed system
supports both quantitative and mode-based reasoning.

\textbf{Conclusion and Future Work.} Combing both Adjoint Logic and
Graded Modal Logic results in a very expressive system capable of
mixing several different notions of substructural logics.  We are
currently proving cut elimination for the sequent calculus, and
substitution, subject reduction, and strong normalization for the term
assignment.  Furthermore, we are currently developing a categorical
model of the system extending the work of
Katsumata~\cite{Katsumata:2018}, and developing an implementation as
an extension of the Granule programming
language~\cite{Orchard:2019}. Granule currently provides a combination
of linear, indexed, and graded modal types in a standard functional
setting. It employs a bidirectional type checking algorithm which
generates complex constraints involving type indices and grades, which
are then passed to an SMT solver. Extending Granule's core to the
pointed semiring approach requires new encodings into the underlying
SMT format. Furthermore, a suitable surface language is in
development, providing access to the two forms of judgment in a
natural style.



\bibliographystyle{eptcs}

\appendix

\section{Full Systems}
\label{sec:full_systems}

\subsection{Sequent Calculus}
\label{subsec:sequent_calculus}
\begin{mdframed}
  \begin{center}
    {\underline{Valid Vector Multiplication}}
    \begin{mathpar}
      \ottdruleVMXXXEmpty{} \and
      \ottdruleVMXXXExt{}
    \end{mathpar}
  \end{center}
\end{mdframed}

\begin{mdframed}
  \begin{center}
    {\underline{Graded Fragment}}
    \begin{mathpar}
      \ottdruleSGXXid{} \and
      \ottdruleSGXXUnitR{} \and
      \ottdruleSGXXTenL{} \and
      \ottdruleSGXXTenR{} \and
      \ottdruleSGXXGrdR{} \and
      \ottdruleSGXXCut{} \and
      \ottdruleSGXXSub{} \and
      \ottdruleSGXXWeak{} \and
      \ottdruleSGXXCont{} \and      
      \ottdruleSGXXEx{}       
    \end{mathpar}
  \end{center}
\end{mdframed}
\vspace{-10pt}
\begin{mdframed}
  \begin{center}
    {\underline{Mixed Fragment} }
    \begin{mathpar}
      \ottdruleSMXXId{} \and
      \ottdruleSMXXUnitR{} \and
      \ottdruleSMXXGTenL{} \and
      \ottdruleSMXXTenL{} \and
      \ottdruleSMXXTenR{} \and
      \ottdruleSMXXImpL{} \and
      \ottdruleSMXXImpR{} \and
      \ottdruleSMXXGrdL{} \and
      \ottdruleSMXXLinL{} \and
      \ottdruleSMXXLinR{} \and
      \ottdruleSMXXCut{} \and
      \ottdruleSMXXGCut{} \and      
      \ottdruleSMXXGSub{} \and
      \ottdruleSMXXGWeak{} \and
      \ottdruleSMXXGCont{} \and      
      \ottdruleSMXXGEx{} \and
      \ottdruleSMXXEx{}       
    \end{mathpar}      
  \end{center}
\end{mdframed}


\subsection{Natural Deduction}
\label{subsec:natural_deduction}
\begin{mdframed}
  \begin{center}
    {\underline{Valid Vector Multiplication}}
    \begin{mathpar}
      \ottdruleVMXXXEmpty{} \and
      \ottdruleVMXXXExt{} 
    \end{mathpar}
  \end{center}
\end{mdframed}

\begin{mdframed}
  \begin{center}
    {\underline{Graded Fragment}}
    \begin{mathpar}
      \ottdruleNGXXId{} \and
      \ottdruleNGXXUnitI{} \and
      \ottdruleNGXXTenI{} \and
      \ottdruleNGXXTenE{} \and
      \ottdruleNGXXGrdI{} \and
      \ottdruleNGXXSub{} \and
      \ottdruleNGXXWeak{} \and
      \ottdruleNGXXCont{} \and      
      \ottdruleNGXXEx{}       
    \end{mathpar}
  \end{center}
\end{mdframed}
\vspace{-10pt}
\begin{mdframed}
  \begin{center}
    {\underline{Mixed Fragment} }
    \begin{mathpar}
      \ottdruleNMXXId{} \and
      \ottdruleNMXXUnitI{} \and
      \ottdruleNMXXTenI{} \and
      \ottdruleNMXXTenE{} \and
      \ottdruleNMXXImpI{} \and
      \ottdruleNMXXImpE{} \and
      \ottdruleNMXXLinI{} \and
      \ottdruleNMXXLinE{} \and
      \ottdruleNMXXGrdE{} \and
      \ottdruleNMXXGSub{} \and
      \ottdruleNMXXGWeak{} \and
      \ottdruleNMXXGCont{} \and
      \ottdruleNMXXGEx{} \and
      \ottdruleNMXXEx{}       
    \end{mathpar}      
  \end{center}
\end{mdframed}


\subsection{Term Assignment}
\label{subsec:term_assignment}
\begin{mdframed}
  \begin{center}
    {\underline{Valid Vector Multiplication}}
    \begin{mathpar}
      \ottdruleVMXXXEmpty{} \and
      \ottdruleVMXXXExt{} 
    \end{mathpar}
  \end{center}
\end{mdframed}

\begin{mdframed}
  \begin{center}
    {\underline{Graded Fragment}}
    \begin{mathpar}
      \ottdruleTGXXId{} \and
      \ottdruleTGXXUnitI{} \and
      \ottdruleTGXXTenI{} \and
      \ottdruleTGXXTenE{} \and
      \ottdruleTGXXGrdI{} \and
      \ottdruleTGXXSub{} \and
      \ottdruleTGXXWeak{} \and
      \ottdruleTGXXCont{} \and      
      \ottdruleTGXXEx{}       
    \end{mathpar}
  \end{center}
\end{mdframed}
\vspace{-10pt}
\begin{mdframed}
  \begin{center}
    {\underline{Mixed Fragment} }
    \begin{mathpar}
      \ottdruleTMXXId{} \and
      \ottdruleTMXXUnitI{} \and
      \ottdruleTMXXTenI{} \and
      \ottdruleTMXXTenE{} \and
      \ottdruleTMXXImpI{} \and
      \ottdruleTMXXImpE{} \and
      \ottdruleTMXXLinI{} \and
      \ottdruleTMXXLinE{} \and
      \ottdruleTMXXGrdE{} \and
      \ottdruleTMXXGSub{} \and
      \ottdruleTMXXGWeak{} \and
      \ottdruleTMXXGCont{} \and
      \ottdruleTMXXGEx{} \and
      \ottdruleTMXXEx{}       
    \end{mathpar}      
  \end{center}
\end{mdframed}



\end{document}